\RequirePackage{fix-cm}
\documentclass[twocolumn]{svjour3}          
\smartqed  

\usepackage[hyphens]{url}
\usepackage[hidelinks]{hyperref}
\hypersetup{breaklinks=true}

\usepackage{lipsum}
\usepackage{balance}
\usepackage{graphicx}
\usepackage[framemethod=TikZ]{mdframed}
\usepackage[misc]{ifsym}
\usepackage{booktabs}
\usepackage{color}
\usepackage{url}
\usepackage{balance}
\usepackage{graphicx}
\usepackage{epstopdf}
\usepackage{array}
\usepackage{amsmath}
\usepackage{amsfonts}
\usepackage{chemarrow}
\usepackage{multirow}
\usepackage{wrapfig}
\usepackage{picinpar}
\usepackage{extarrows}
\usepackage{microtype} 
\usepackage{wrapfig}
\usepackage{amssymb,amsbsy,tabulary}
\usepackage{authblk}
\usepackage{algorithm,algorithmic}
\newcommand{\ALOOP}[1]{\ALC@it\algorithmicloop\ #1%
  \begin{ALC@loop}}
\newcommand{\ENDALOOP}{\end{ALC@loop}\ALC@it\algorithmicendloop}

\authorrunning{Ryu et al.}

\makeatletter
\long\def\figwindownonum[#1,#2,#3,#4] {
\begin{window}[#1,#2,{#3},{\centering#4\par}] }
\def\endfigwindownonum{\end{window}}
\makeatother
\begin{document}

\title{Can Differential Privacy Practically Protect Collaborative Deep Learning Inference for IoT?}

\author{Jihyeon Ryu \and Yifeng Zheng* \and Yansong Gao \and Alsharif Abuadbba \and Junyaup Kim \and Dongho Won \and  Surya Nepal \and Hyoungshick Kim \and Cong Wang  
}


\institute{Jihyeon Ryu, Junyaup Kim, Dongho Won, Hyoungshick Kim  \at
              Department of Computer Science and Engineering, Sungkyunkwan University, South Korea \\
              \email{jhryu@security.re.kr, yaup22cc@likelion.org, dhwon@security.re.kr, hyoung@skku.edu}           
           \and
           Yifeng Zheng (*Corresponding author) \at
              School of Computer Science and Technology, Harbin Institute of Technology, Shenzhen, China\\
             \email{yifeng.zheng@hit.edu.cn}
            \and
            Yansong Gao \at
            School of Computer Science and Engineering, Nanjing University of Science and Technology, Nanjing, China \\
            \email{yansong.gao@njust.edu.cn}
            \and
            Alsharif Abuadbba, Surya Nepal \at
            Data61, CSIRO, Australia\\
            \email{Sharif.Abuadbba@data61.csiro.au, Surya.Nepal@data61.csiro.au}
            \and 
            Cong Wang \at
            Department of Computer Science, City University of Hong Kong, Hong Kong, China\\
            \email{congwang@cityu.edu.hk}
}

\date{Received: date / Accepted: date}

\maketitle

\begin{abstract}
Collaborative inference has recently emerged as an attractive framework for applying deep learning to Internet of Things (IoT) applications by splitting a DNN model into several subpart models among resource-constrained IoT devices and the cloud. However, the  reconstruction attack was proposed recently to recover the original input image from intermediate outputs that can be collected from local models in collaborative inference. For addressing such privacy issues, a promising technique is to adopt differential privacy so that the intermediate outputs are protected with a small accuracy loss. In this paper, we provide the first systematic study to reveal insights regarding the effectiveness of differential privacy for collaborative inference against the reconstruction attack. We specifically explore the privacy-accuracy trade-offs for three collaborative inference models with four datasets (SVHN, GTSRB, STL-10, and CIFAR-10). Our experimental analysis demonstrates that differential privacy can practically be applied to collaborative inference when a dataset has small intra-class variations in appearance. With the (empirically) optimized privacy budget parameter in our study, the differential privacy technique incurs accuracy loss of 0.476\%, 2.066\%, 5.021\%, and 12.454\% on SVHN, GTSRB, STL-10, and CIFAR-10 datasets, respectively, while thwarting the reconstruction attack.

\end{abstract}

\section{Introduction}
Recent advancements in deep learning techniques have greatly empowered various Internet of Things (IoT) applications such as object recognition, human activity recognition, health monitoring, and environmental sensing \cite{YaoHZZA17,RaduTBLMMK17,YaoZSZZLA17,YaoZSZZHLLSA18}.
However, running a trained deep neural network (DNN) model with new inputs (i.e., DNN inference) would be resource-intensive and requires massive computational resources, making it notably difficult to be directly deployed on resource-constrained IoT devices \cite{yao2018deep,YaoZSLLSA18}.
Therefore, an alternative practical way for deployment is to construct a DNN model on a cloud server and forward input data from IoT devices to the cloud server for the inference.
With such deployment, however, IoT devices' data are inevitably exposed to the cloud service provider, raising privacy concerns for some IoT applications that would process sensitive and/or private data.
 
\begin{figure}[t!]
\centerline{\includegraphics[width=0.48\textwidth]{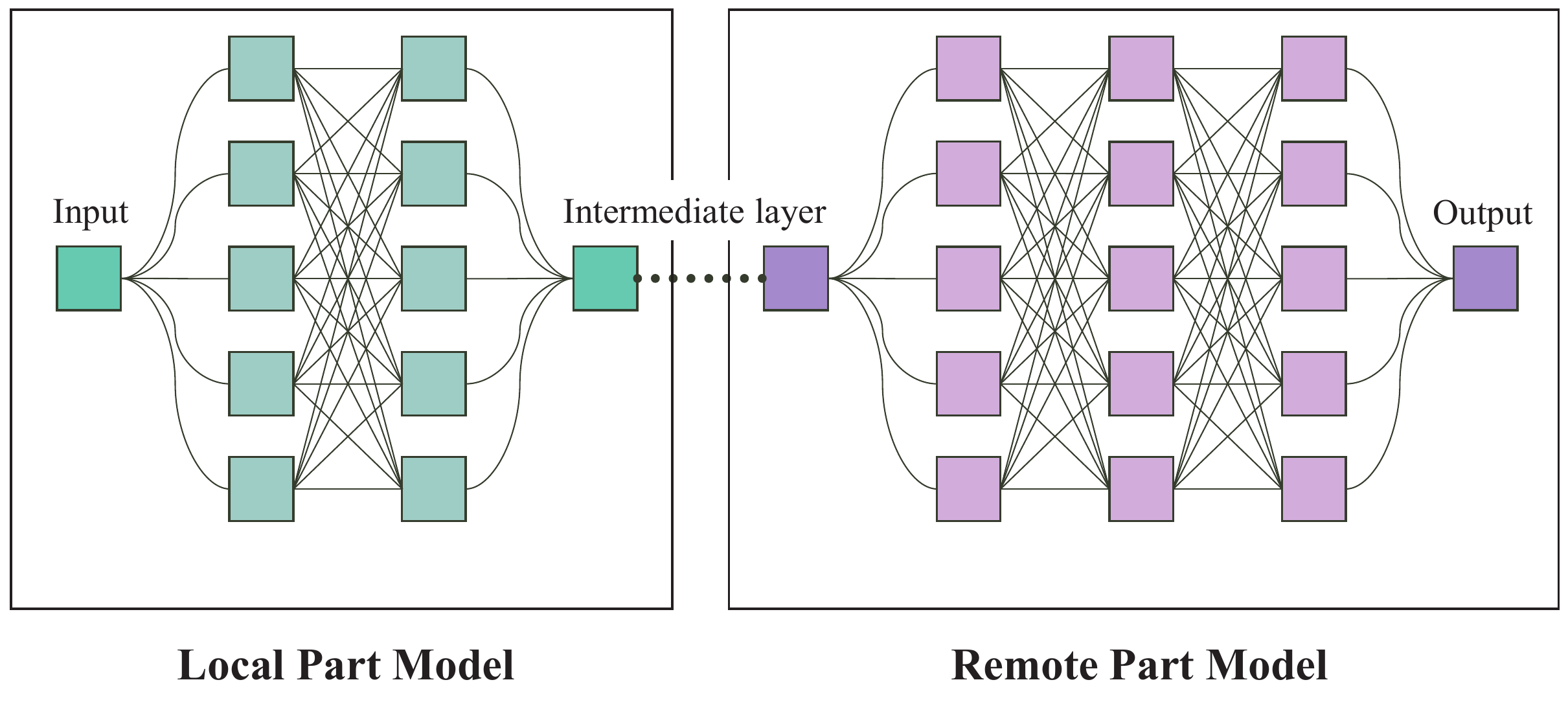}}
\caption{Overview of collaborative inference.}
\label{fig:Flowchart}
\end{figure}

Recently, \emph{collaborative inference}~\cite{Teerapittayanon17,KoNAM18} was introduced to avoid the direct exposure of such data from resource-constrained IoT devices, which DNN inference can still be effectively supported. 
In particular, in the collaborative inference framework, a DNN model is split into a local part model containing simple shallow layers of the DNN model and a remote part model containing the remaining sophisticated layers. The local part model is typically deployed on the resource-constrained IoT devices, while the remote part model is deployed on the cloud, as illustrated in Fig.~\ref{fig:Flowchart}. 

In this collaborative inference framework, DNN inference is performed collaboratively, crossing from the local part model to the remote part model. The local part model first processes input data to obtain an intermediate output. Then, this intermediate output is sent to the remote part model to perform forward inference computation over the remaining layers. Consequently, collaborative inference fundamentally eschews direct exposure of the raw input data to the cloud. Moreover, collaborative inference has clear advantages for reducing the computational resources of IoT devices in deep learning applications.
%
%
In the view of model providers, the use of collaborative inference would be preferred as well because they do not need to give out the entire DNN model for deployment on local devices.

At first glance, it may seem sufficient to use collaborative inference for protecting raw input data used in a DNN model. However, recent studies~\cite{WangZBZCY18,HeZL19} show that collaborative inference could entail privacy risks. The intermediate output produced from the local part model can contain sensitive information used to recover the original raw input data. He \textit{et al.} \cite{HeZL19} presented the feasibility of a reconstruction attack targeting collaborative inference for image-based applications, which is designed for an honest-but-curious cloud service provider to recover the original input image from the intermediate output generated from the local part model. In independent work, to mitigate the privacy risks from exposing the intermediate output, Wang \textit{et al.} \cite{WangZBZCY18} proposed a collaborative inference framework using \emph{differential privacy} \cite{Dwork06} to avoid the privacy leakage from the intermediate output. Differential privacy has become the de facto privacy standard as it provides a rigorous mathematical framework for formalizing privacy guarantees in terms of the privacy budget $\epsilon$. The framework in \cite{WangZBZCY18} employs differential privacy via adding delicately calibrated noises to the intermediate output values. As such perturbations definitively incur a degradation on the inference accuracy, the framework further delicately provides a noisy training technique to endow the remote part model with robustness to perturbed data and alleviate the impact of noise perturbation on the inference accuracy.

To deploy a collaborative inference framework using differential privacy in the real world, it would be necessary for a given collaborative inference model to show that a reasonable budget $\epsilon$ for differential privacy can be chosen against such data reconstruction attacks. However, Wang \textit{et al.} \cite{WangZBZCY18} did not thoroughly analyze the privacy-accuracy trade-offs in the presence of the state-of-the-art data reconstruction attack against collaborative inference. Therefore, our work was motivated by the following research question:

\vspace{0.3cm}
\begin{mdframed}[backgroundcolor=black!10,rightline=false,leftline=false,topline=false,bottomline=false,roundcorner=2mm]
   \textbf{RQ:} Is it feasible to adopt differential privacy to gain protection in collaborative inference against reconstruction attack while preserving high accuracy of the inference? 
\end{mdframed}
\vspace{0.3cm}

To answer the research question, we implement the state-of-the-art differential privacy framework for collaborative inference~\cite{WangZBZCY18}, and newly apply the state-of-the-art reconstruction attack against that framework over various datasets to reveal the privacy-accuracy trade-offs. To our best knowledge, our study is the first that assesses the practical usability of differential privacy for collaborative inference in the presence of the state-of-the-art reconstruction attack.
We summarize the key contributions of this paper as follows: 

\begin{itemize}

\item We implement the state-of-the-art collaborative inference framework using differential privacy~\cite{HeZL19,WangZBZCY18} and reconstruction attack to analyze the privacy-accuracy trade-offs in collaborative inference. 

\item We conduct extensive evaluations on the attack and defense implementations with various datasets, including SVHN, GTSRB, STL-10, and CIFAR-10, and varying the privacy budget $\epsilon$. Unlike the previous study with a fixed split~\cite{WangZBZCY18}, we evaluate several split settings by varying the layers of the local part model and the remote part model. We find that the effectiveness of differential privacy increases as the number of layers of the local part model decreases in the collaborative inference model deployment (i.e., the number of layers of the remote part model increases). 

\item We reveal insights about how the effectiveness of differential privacy is significantly affected by the characteristics of datasets through our experiments. We find that differential privacy would be more effective when a dataset has small intra-class variations in appearance. In our experiments, the best privacy budget $\epsilon$ incurs accuracy loss of 0.476\%, 2.066\%, 5.021\%, and 12.454\% on SVHN, GTSRB, STL-10, and CIFAR-10 datasets, respectively, while preventing data reconstruction attacks.
\end{itemize}

The remainder of this paper is organized as follows.
Section \ref{sec:preliminaries} provides background information on the differential privacy-based collaborative inference framework and the data reconstruction attack.
Section \ref{sec:evaluation_study} presents comprehensive experimental evaluations.
Section \ref{sec:discussion} discusses key findings from our extensive evaluations and draws practical insights.
Section \ref{sec:related_work} describes the related work.
Section \ref{sec:conclusion} concludes this work.

\section{Background}
\label{sec:preliminaries}

\subsection{Collaborative Inference}

In the collaborative inference framework \cite{Teerapittayanon17,KoNAM18} for IoT-cloud applications, as shown in Fig. \ref{fig:Flowchart}, a trained DNN model, denoted by $f_{\theta}$ and parameterized by model parameters $\theta$, is split into two parts: a local part model $f_{\theta_1}$ and a remote part model $f_{\theta_2}$. The former is deployed on the client-side (resource-limited IoT devices), while the latter is on the cloud side.
To perform inference for a data sample $\mathbf{x}$, the client first feeds $\mathbf{x}$ to the local part model and obtains $\mathbf{x}^*=f_{\theta_1}(\mathbf{x})$, which represents an intermediate output. 
This intermediate output is then sent to the cloud, which further applies the remote part model $f_{\theta_2}$ to $\mathbf{x}^*$ and produces the ultimate inference result $y=f_{\theta_2}(\mathbf{x}^*)$.

\subsection{Differential Privacy}

Differential privacy is a mathematical framework defined for privacy-preserving data analysis.
The formal definition of $\epsilon$-differential privacy is as follows \cite{Dwork06}.

\begin{definition}
Given any two neighboring inputs $D$ and $D'$ which differ in only one data item, a mechanism $\mathcal{M}$ provides $\epsilon$-differential privacy if $Pr[\mathcal{M}(D)\in S]\leq e^{\epsilon}\cdot Pr[\mathcal{M}(D')\in S]$.  
\end{definition}

Intuitively, the above definition indicates that for any output in the range $S$ of the mechanism $\mathcal{M}$, its probability of being produced from $D$ is very close to that of being produced from $D'$, as characterized by $\epsilon$.
That is, given any output, one can hardly tell whether it is produced from $D$ or $D'$.
The parameter $\epsilon$ is usually referred to as the privacy budget.
A smaller $\epsilon$ value indicates \textit{stronger} privacy protection.

To achieve differential privacy, the common approach is to add calibrated noises to the output of a function $g(\cdot)$ based on specific probability distributions \cite{DworkMNS06}.
A widely used probability distribution in differential privacy is the Laplace distribution, denoted by $Lap(b)$, where $b$ is called the scale parameter.
In particular, the probability density function is: $Pr[x]=\frac{1}{2b}e^{-|x|/b}$.
The Laplace mechanism \cite{DworkMNS06} for $\epsilon$-differential privacy works by sampling noises from $Lap(b)$ and adding the noises to the output values of the function $g(\cdot)$.
Here, to achieve $\epsilon$-differential privacy, $b$ is set according to the global sensitivity $\Delta g$ of the function $g(\cdot)$, i.e., $b=\frac{\Delta g}{\epsilon}$.
Let $||\cdot||_1$ denote the $l_1$ norm.
The global sensitivity of $\Delta g$ is defined as:
\[\Delta g= \max\limits_{D,D'} ||g(D)-g(D')||_1. \]

\begin{algorithm}[!t]
\caption{Differential Privacy Scheme Using Local Perturbation}
\label{algo:local_perturbation}
\begin{algorithmic}[1]

\REQUIRE Input data sample $\mathbf{x}$; Bound threshold $B$; Privacy budget $\epsilon$.

\ENSURE Noisy intermediate output $\mathbf{x}^*$.

\STATE $\tilde{\mathbf{x}} \leftarrow f_{\theta_1}(\mathbf{x})$

\STATE $d=||\tilde{\mathbf{x}}||_{\infty}$
\STATE $\tilde{\mathbf{x}} \leftarrow \tilde{\mathbf{x}}/ max(1, \frac{d}{B} ) $
\STATE $\mathbf{x}^* \leftarrow \tilde{\mathbf{x}} +Lap(2B/\epsilon)$ // \small{Sampling and adding noises element-wise.}
\end{algorithmic}
\end{algorithm}

\subsection{Differential Privacy for Collaborative Inference}

The differential privacy framework for the collaborative inference that we investigate herein is the state-of-the-art by Wang \textit{et al.} \cite{WangZBZCY18}.
At a high level, this framework is comprised of two modules: one module on the local part which performs the differential privacy noise-based perturbation in the inference phase; and the other module on the remote part, which conducts a noisy training process to mitigate the impact of noise perturbation on the inference accuracy performance of a DNN model.

As shown in Algorithm \ref{algo:local_perturbation}, the differential privacy based noise perturbation proceeds as follows.
Given an input data sample $x$, the client passes it to the local part model and obtains $\tilde{\mathbf{x}}\leftarrow f_{\theta_1}(\mathbf{x})$.
Then, noises sampled from the Laplace distribution are added to a bounded version of $\tilde{\mathbf{x}}$, producing the noisy intermediate output $\mathbf{x}^*$, which is sent to the cloud for inference.
Note that bounding each value in $\tilde{\mathbf{x}}$ is needed because it is hard to directly estimate the global sensitivity of $f_{\theta_1}(\mathbf{x})$ for adding differential privacy noises.
The bound $B$, as used in Algorithm \ref{algo:local_perturbation}, can be set as the median of the infinity norm with regard to a set of training examples during the training phase.
We note that the client could optionally perform nullification on the input data sample $\mathbf{x}$ by randomly setting a portion of elements in $\mathbf{x}$ to zeros, masking some parts of $\mathbf{x}$ that are deemed highly sensitive.

As the perturbation will obviously degrade the accuracy performance of the DNN model, the design \cite{WangZBZCY18} constructively takes advantage of noisy training to fine-tune the remote part model $f_{\theta_2}(\cdot)$.
The main idea is to perform training on both plain representations and noisy counterparts for the remote part model, taking into account the training losses for both plain representations and noisy representations.
Here, a clear representation means the intermediate output obtained by passing an input data sample to the original clean local part model.
We refer interested readers \cite{WangZBZCY18} for details on the algorithm for noisy training.
Let $f'_{\theta_2}(\cdot)$ denote the fine-tuned remote part model.
In the inference phase, upon receiving the noisy intermediate output from the client, the cloud conducts the inference by passing it to $f'_{\theta_2}(\cdot)$ and returns $f'_{\theta_2}(\mathbf{x}^*)$ to the client as the inference result.

\noindent \textbf{Remark.} It is noted that since our goal is to evaluate the practical usability of the above state-of-the-art framework in the presence of the reconstruction attack, we exactly follow the Laplace mechanism-based construction in \cite{WangZBZCY18} and do not aim to propose new differential privacy mechanisms that can work for collaborative inference.
We are aware that there are other mechanisms like the Gaussian mechanism and the exponential mechanism \cite{BaiLLYJX22}.
However, we emphasize that whether and how they can be effectively applied to the collaborative inference paradigm remains unclear.
Indeed, it is non-trivial to apply differential privacy to the collaborative inference paradigm because simply adding noises locally will lead to poor utility of the inference service.
This also accounts for why the prior work \cite{WangZBZCY18} needs to develop an algorithm for fine-tuning the model training process at the cloud serve, so as to balance privacy and utility.
If there emerge other custom and workable differential privacy mechanisms for collaborative inference later, it would be interesting and valuable as well to explore their effectiveness against the reconstruction attack. In that case, we believe our initial study in this area can serve as good pointers and references.

\begin{algorithm}[!t]
\caption{Reconstruction Attack}
\label{algo:reconstruction-attack}
\begin{algorithmic}[1]

\REQUIRE local part model $f_{\theta_1}$; Intermediate output $f_{\theta_1}(x_0)$ for input $x_0$; Maximum number of iterations $T$; Hyperparameters $\lambda$ for total variation and $s$ for step size.

\ENSURE Reconstructed $\widehat{x}$ for $x_0$

\STATE $L(x)=||f_{\theta_1}(x)-f_{\theta_1}(x_0)||^2_2+\lambda\cdot TV(x)$

\STATE $t=0$

\STATE $x^{(0)}=\mathsf{Init}()$

\STATE \textbf{While} $(t<T)$ \textbf{do}
\STATE $~~~~~~~~~~~~~~~~~~x^{(t+1)}=x^{(t)}-s\cdot  \frac{\partial L(x^{(t)})}{\partial x^{(t)}}$
\STATE $~~~~~~t=t+1$
\STATE \textbf{end}

\STATE return $\widehat{x}=x^{(T)}$

\end{algorithmic}
\end{algorithm}

\subsection{Reconstruction Attack against Collaborative Inference}
\label{reconstructionattack}

In a recent work \cite{HeZL19}, He \textit{et al.} proposed reconstruction attacks that allow the cloud to reconstruct the input image given the intermediate output and the local part model in the collaborative inference framework.
Our study focuses on the reconstruction attack in the white-box setting because it is much stronger than that in the black-box setting.
Evaluating differential privacy in the most powerful white-box attack setting arguably can better reflect how useful differential privacy can be in practice.
For this attack setting, the local part model is known to the cloud, given that the whole DNN model is trained by the cloud, which also performs model splitting and provides the local part to the client. 
It is noted that the attack is proposed against images, so our evaluations are performed over image datasets.
For other data types, we are not aware of any works that propose corresponding reconstruction attacks in the context of collaborative inference.
Meanwhile, we note that the evaluation in the prior work \cite{WangZBZCY18} designing the differential privacy framework for collaborative inference is also dominated by image datasets. 
Further, it is worth noting that one main motivation for the collaborative inference paradigm initially proposed in \cite{Teerapittayanon17} is to allow the local client to send to the cloud server a much smaller intermediate output rather than the large-sized raw input, for which image data as the input will benefit the most from such paradigm. Indeed in the seminal work \cite{Teerapittayanon17}, the evaluation is also conducted over image datasets.

Algorithm \ref{algo:reconstruction-attack} gives the details of the studied reconstruction attack that aims to reconstruct input images in collaborative inference. 
Let $x_0$ denote an example input image and $\widehat{x}$ denote the reconstructed image against $x_0$.
The main idea is to formulate the reconstruction attack as an optimization problem under two requirements.
Firstly, feeding $\widehat{x}$ to the local part model $f_{\theta_1}$ produces an intermediate output $f_{\theta_1}(\widehat{x})$ that is similar to the observed $f_{\theta_1}(x_0)$. Here the similarity is measured by the Euclidean distance. Secondly, $\widehat{x}$ is a natural image which follows the same distribution as the input samples for the DNN model.
For this requirement, the total variation measure is adopted to enforce that the reconstructed image $\widehat{x}$ is as piece-wise smooth as possible.

\section{Comprehensive Evaluations}
\label{sec:evaluation_study}

\begin{table}[t!]
\caption{Specifications of Datasets and Clipping Bounds under Different Splitting Cases}
\centering
\resizebox{0.48\textwidth}{!}{
\begin{tabular}{@{}ccccc@{}}
\toprule
Dataset  & SVHN & GTSRB & STL-10 & CIFAR-10 \\ \midrule 
Training Set Size & 73,200 & 14,600 & 10,000 & 50,000 \\
Testing Set Size & 26,000 & 4,800 & 3,000 & 10,000 \\
Case 1 Bound & 250.104 & 237.374 & 230.409 & 230.174 \\
Case 2 Bound & 7477.173 & 16385.918 & 12801.195 & 13058.996 \\
Case 3 Bound & 7774.149 & 9613.522 & 8346.818 & 10680.272 \\
\bottomrule
\end{tabular}
}
\label{table:dataset_info}
\end{table}

\subsection{Experimental Setup}

\begin{figure}[t!]
\centerline{\includegraphics[width=0.48\textwidth]{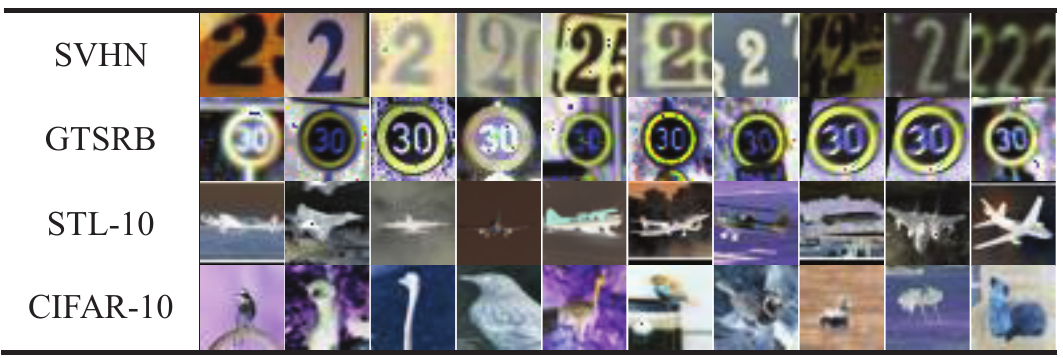}}
\caption{Intra-class variation of each dataset: SVHN (digit number 2), GTSRB (30 km/h speed limit signs), CIFAR10 (bird), and STL10 (airplane). It is observable that SVHN $<$ GTSRB $<$ STL-10 $<$ CIFAR-10 in terms of intra-class variation degree.}
\label{fig:ONEclasses}
\vspace{-10pt}
\end{figure}

\noindent\textbf{Datasets.} 
We use four datasets in our comprehensive empirical evaluations, including SVHN \cite{svhn}, GTSRB \cite{gtsrb}, CIFAR-10 \cite{cifar10}, and STL-10 \cite{stl10}. Fig. \ref{fig:ONEclasses} show the one class of each dataset. The overall specifications of these datasets are given in Table \ref{table:dataset_info}. It is noted that for each dataset, the clipping bound as shown in Table \ref{table:dataset_info} is derived by computing the median of the infinity norms of intermediate outputs with regard to $100$ randomly chosen training examples. Each dataset is introduced in more details below:

\begin{itemize}
    \item SVHN. This dataset contains labeled images of house numbers in Google Street View images. Each image has a size of $(32,32,3)$, and is labeled from $0$ to $9$. We randomly select $73,200$ images for training and $26,000$ for testing. 
    
    \item GTSRB. This dataset contains labeled images of traffic sign images. The images have 3 channels but with varying sizes, and are categorized into more than 40 classes. There are more than $50,000$ images in total. In our evaluation, we randomly select $14,600$ images out of 10 classes for training and $4,800$ images for testing, with each image being resized to $(32,32,3)$. 
    
    \item STL-10. This dataset contains labeled images of natural objects in 10 classes. There are $1,300$ images in each class. Each image has a size of $(96, 96, 3)$. We randomly select $10,000$ images for training and $3,000$ images for testing, with each image being resized to $(32,32,3)$. 
    
    \item CIFAR-10. This dataset also contains labeled images of natural objects in 10 classes (such as airplane, bird, car, and cat), with $6,000$ images per class. Each image has a size of $(32, 32, 3)$. There are $50,000$ training images and $10,000$ testing images, which are used in our evaluation.
        
\end{itemize}

\begin{figure*}[t!]
\centerline{\includegraphics[width=0.8\textwidth]{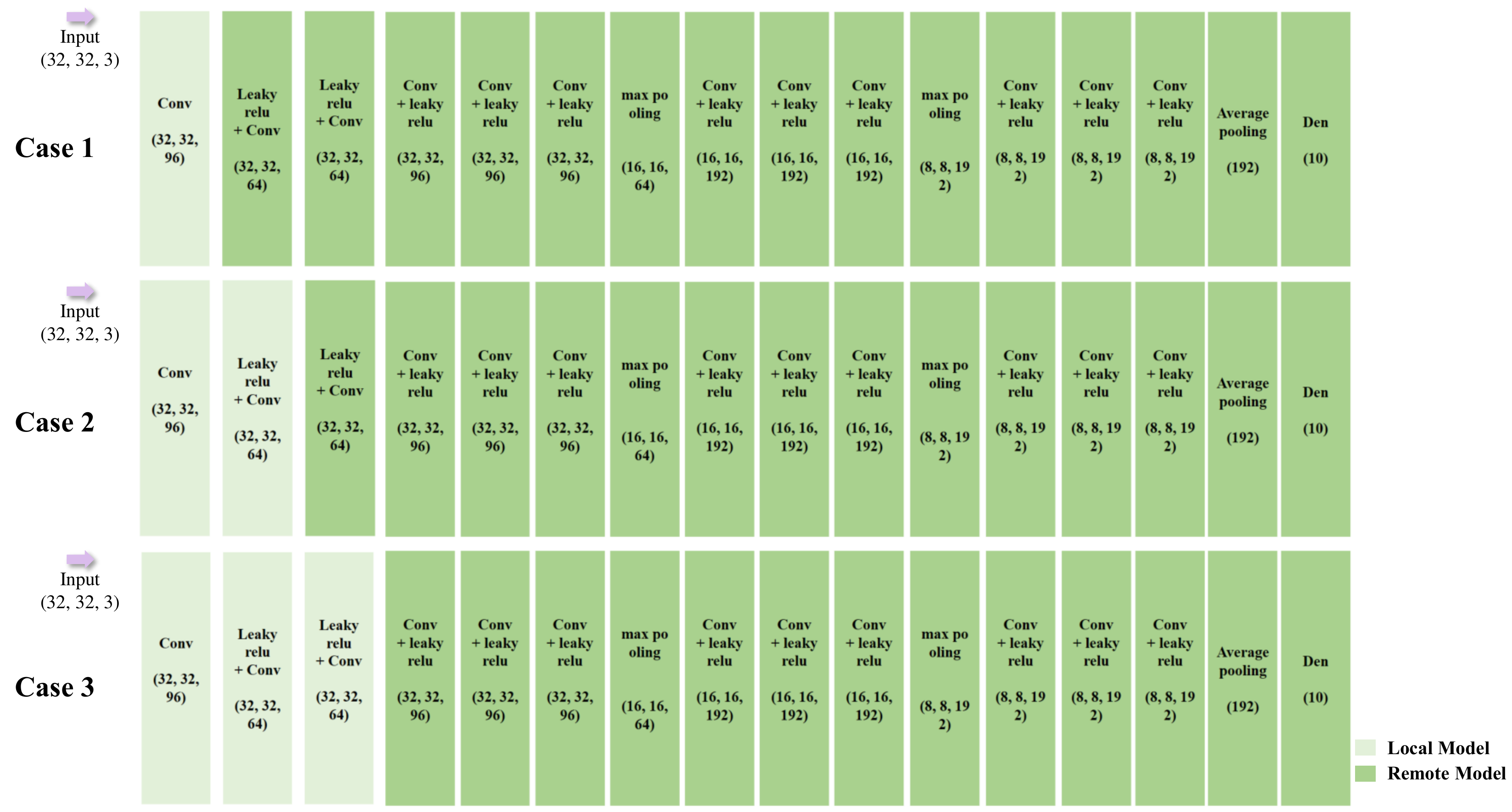}}
\caption{The DNN architecture in our evaluation with varying splitting cases.}
\label{fig:DNN-model}
\end{figure*}

\noindent\textbf{Neural Network Architectures.} The overall DNN architecture used in our evaluation is detailed in 
Figure \ref{fig:DNN-model}.
Case 3 is the same as in \cite{WangZBZCY18}\footnote{Batch normalization is applied in our case to further improve the plain model accuracy.} (where the local model contains 3 convolutional layers).
We have considered more splitting cases: In Case 1, the local part model contains one convolutional layer; In Case 2, the number of local convolutional layers is 2.
%
%
%
More details are given in Figure \ref{fig:DNN-model}.

The input size is $(32, 32, 3)$, and the number of output class is 10.
Following the prior work \cite{WangZBZCY18}, we first derive the model parameters of the local part model (in different cases) from a pre-trained model over CIFAR-100 dataset, and then keep the local part model frozen for the client.
%
That is, the local part model serves as a \textit{generic feature extractor and is applicable to all different datasets} \cite{WangZBZCY18}.
%
%
We trained the remote part model in a fine-tuned manner per each dataset which is introduced above (SVHN, GTSRB, STL-10, and CIFAR-10).
Note that the input for the remote part model is the output obtained by feeding the data sample to the local part model.
%

\vspace{0.2cm}
\noindent\textbf{Hyperparameters.}
For each dataset, we use the ADAM optimizer for training of the remote part model, following \cite{WangZBZCY18}. In order to determine the hyper-parameters, we follow the scale of the hyper-parameters in the prior work \cite{WangZBZCY18} as starting points, and then further fine-tune the hyper-parameters during our training process. 
The learning rate is set to 0.00001 for SVHN, 0.000002 for GTSRB, 0.0000027 for STL-10, and 0.00001 for CIFAR-10, respectively.
The batch size being used is 300 for SVHN, 200 for GTSRB, 200 for STL-10, and 100 for CIFAR-10,  respectively.
The number of training epochs is 40 for SVHN, 100 for GTSRB, 500 for STL-10, and 100 for CIFAR-10, respectively.
Similar to prior work related with the evaluation of differential privacy in other contexts \cite{Jayaraman019}, we vary the privacy budget $\epsilon$ between $0.1$ and $5000$ which represents a wide range, and evaluate the results on accuracy and privacy strengths in the presence of the reconstruction attack. 
It is noted that the presented accuracy results are averaged over 5 runs.

\vspace{0.2cm}
\noindent \textbf{Quantitative Metrics.}
%
In addition to the visualization of reconstructed images, MSE, SSIM, and PSNR metrics are also adopted to quantify the reconstruction efficacy, which generally measures the difference between the original image and the reconstructed image.
    Let $A$ and $B$ denote the original image and reconstructed image respectively, with the size of $m\times n$.
    The pixel value at position $(i,j)$ is denoted by $A(i,j)$ and $B(i,j)$ respectively for images $A$ and $B$.
In what follows we introduce each metric:

\begin{enumerate}
    \item {\it Mean Squared Error} (MSE) measures the similarity between two images by computing the cumulative squared error of pixel values. The lower the value of MSE, the higher the similarity between two images. Specifically, it is computed via:
\[MSE(A, B)= \frac{1}{m\cdot n}\sum_{i, j=1, 1}^{m, n} ||A(i, j) - B(i, j)||^{2}. \]

\item {\it Structural similarity} (SSIM)~\cite{wang2004image} is a perception-based metric which measures the similarity between two images based on structural information. It is computed as:

%
\[SSIM(A, B)= \frac{(2\mu_{A}\mu_{B} + C_1)(2\sigma_{AB} + C_2)}{ (\mu_{A}^{2} + \mu_{B}^2 + C_1)(\sigma_{A}^{2} + \sigma_{B}^2 + C_2)}, \]
where $\mu_A$ and $\mu_B$ are the mean value of pixels in image $A$ and $B$, $\sigma_A^2$ and $\sigma_B^2$ are the variances, and $\sigma_{AB}$ is the co-variance, respectively. In addition, $C_1$ and $C_2$ are constants. The value of SSIM lies between the range of $[0,1]$, and a larger SSIM value indicates a higher similarity between two images.

\item {\it Peak signal-to-noise ratio} (PSNR) measures the similarity of two images via the peak error. Larger PSNR values indicate higher image similarity. It is computed via:
\[PSNR(A, B) = 10\log_{10}(\frac{255^2}{MSE(A, B)}). \]
\end{enumerate}

\subsection{Results over the SVHN Dataset}

\begin{figure}[t!]
\centering
   \begin{minipage}[t]{0.48\linewidth}
      \centering{\includegraphics[width=\linewidth]{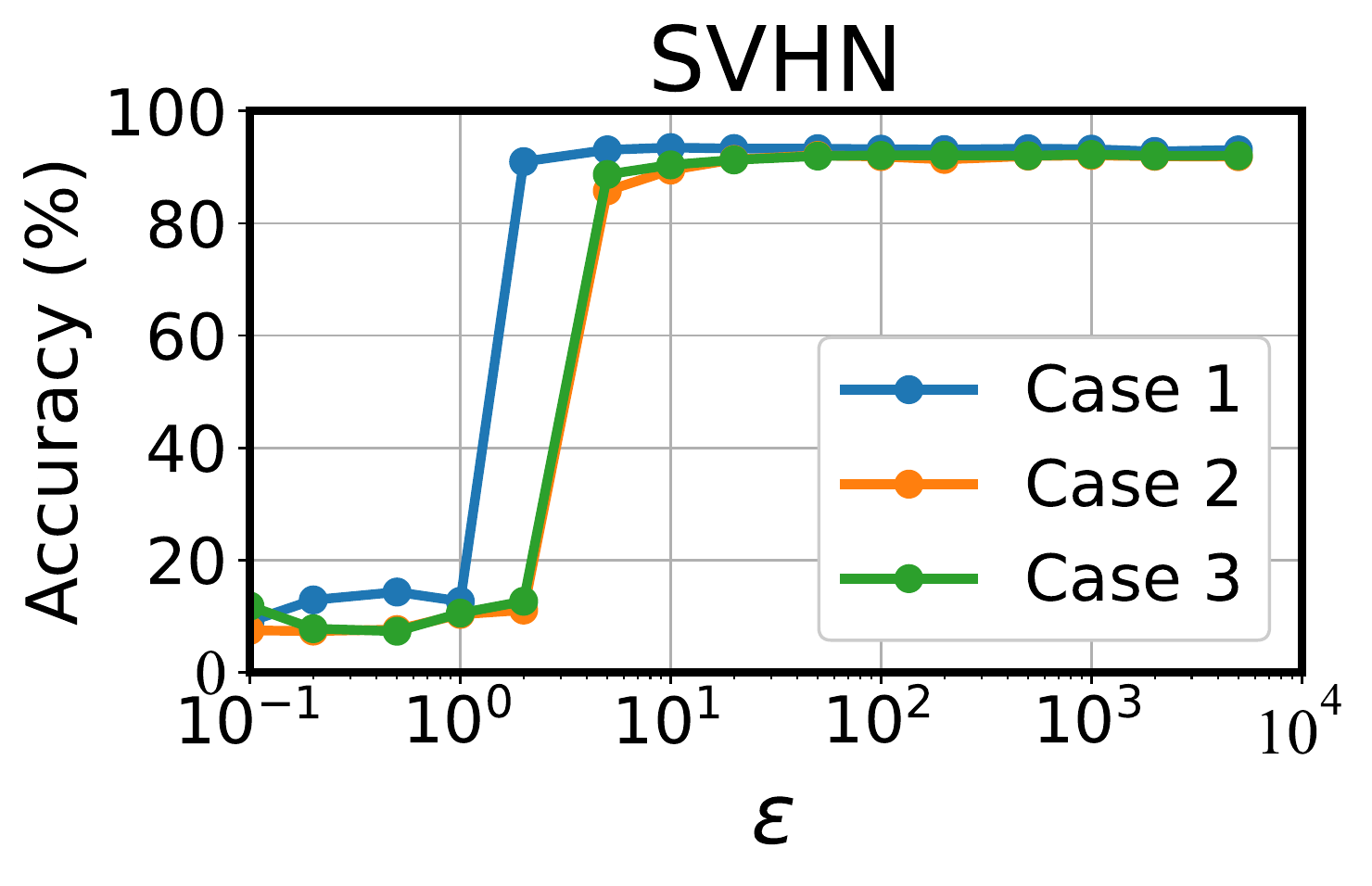}}\\ (a)
   \end{minipage}%
   \begin{minipage}[t]{0.47\linewidth}
      \centering{\includegraphics[width=\linewidth]{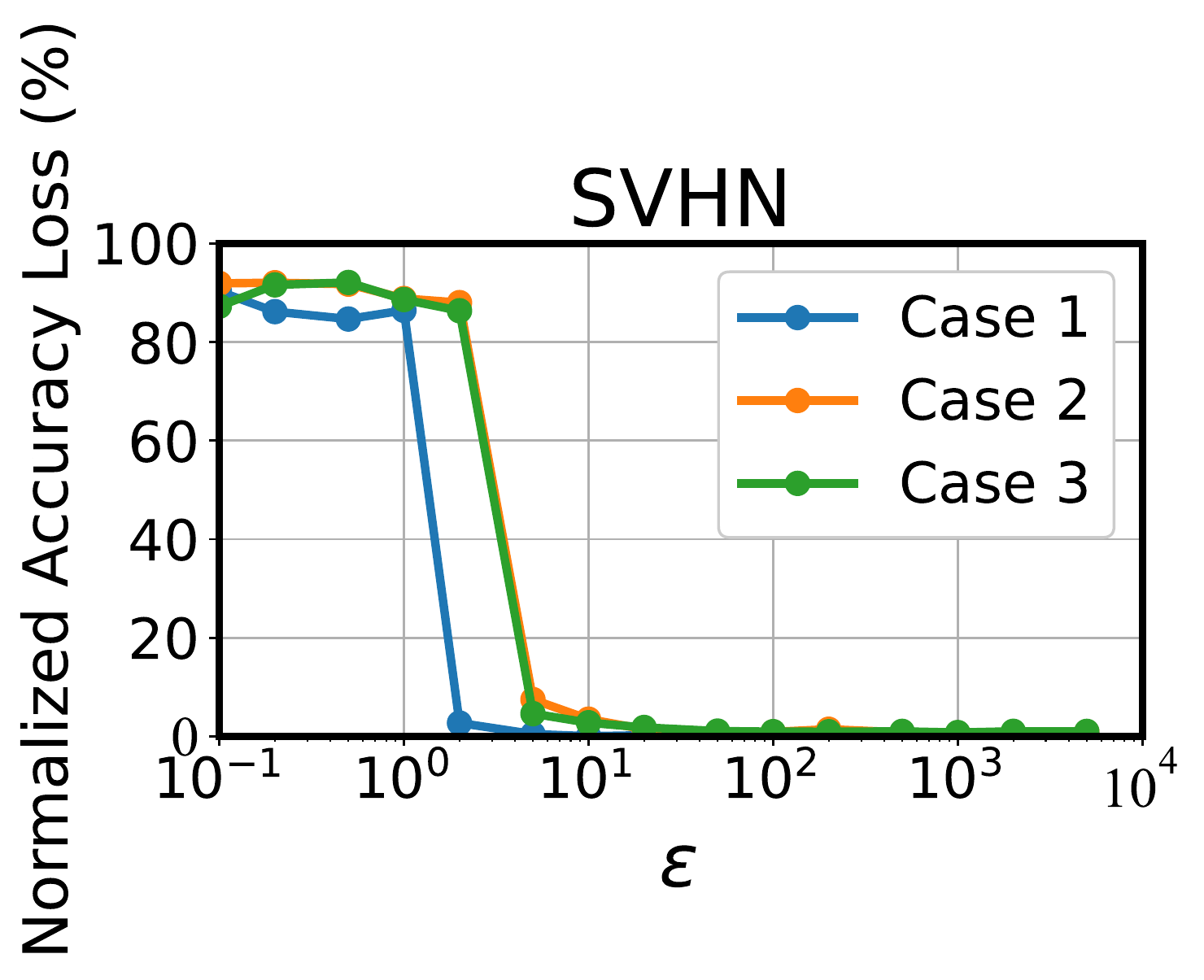}}\\ (b)
   \end{minipage}
\caption{Impact of the privacy budget $\epsilon$ on accuracy (SVHN).}
\label{fig:svhn_acc}

\end{figure}

\begin{figure*}[t!]
\centerline{\includegraphics[width=0.95\textwidth]{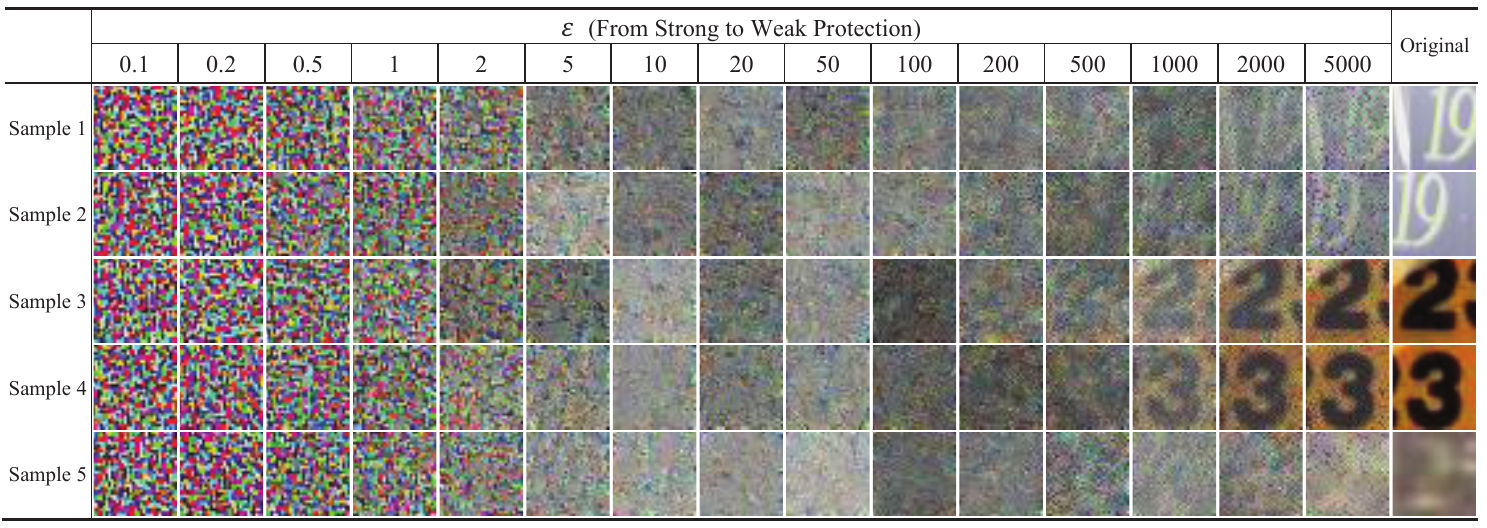}}
\caption{Visual results of applying the attack against the DP method (SVHN Case 3).}
\label{fig:SVHNattackedimage}
\end{figure*}

\begin{figure*}[t!]
\centerline{\includegraphics[width=1\textwidth]{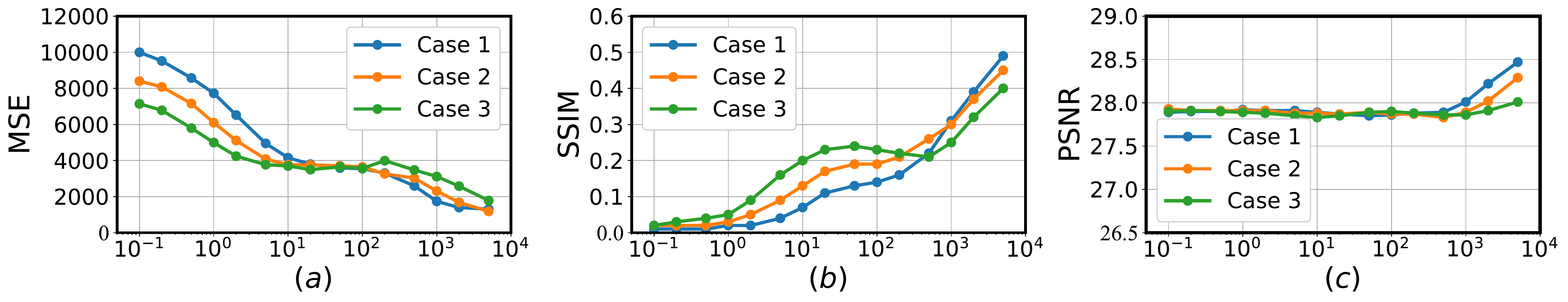}}
\caption{Evaluation of the quantitative metrics for the reconstruction attack efficacy (SVHN): (a) MSE; (b) SSIM; (c) PSNR. }
\label{fig:quantified_results_svhn}
\end{figure*}

\begin{figure}[!t]
\centering
   \begin{minipage}[t]{0.48\linewidth}
         \centering{\includegraphics[width=\linewidth]{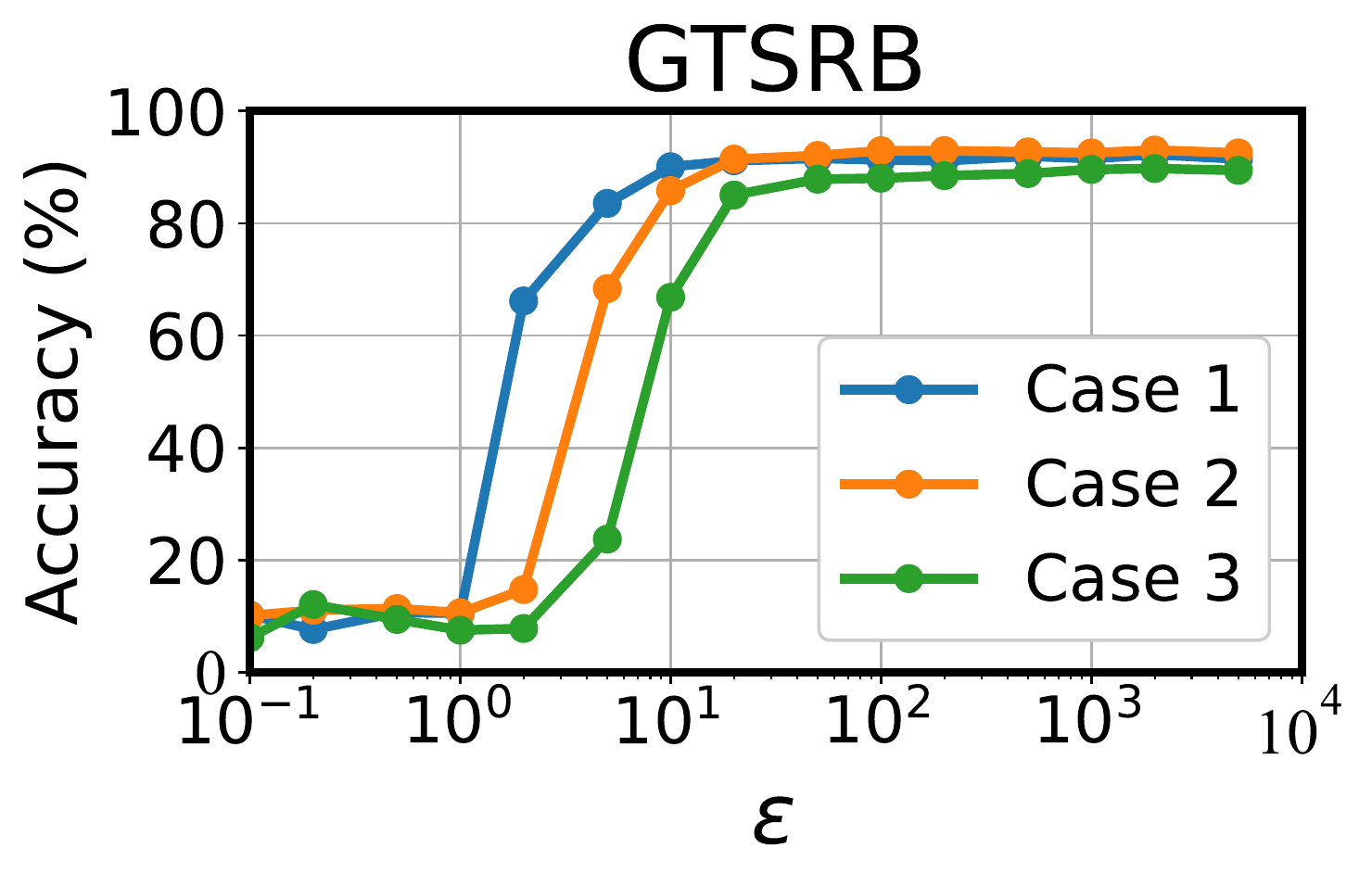}}\\ (a)
   \end{minipage}%
   \begin{minipage}[t]{0.47\linewidth}
      \centering{\includegraphics[width=\linewidth]{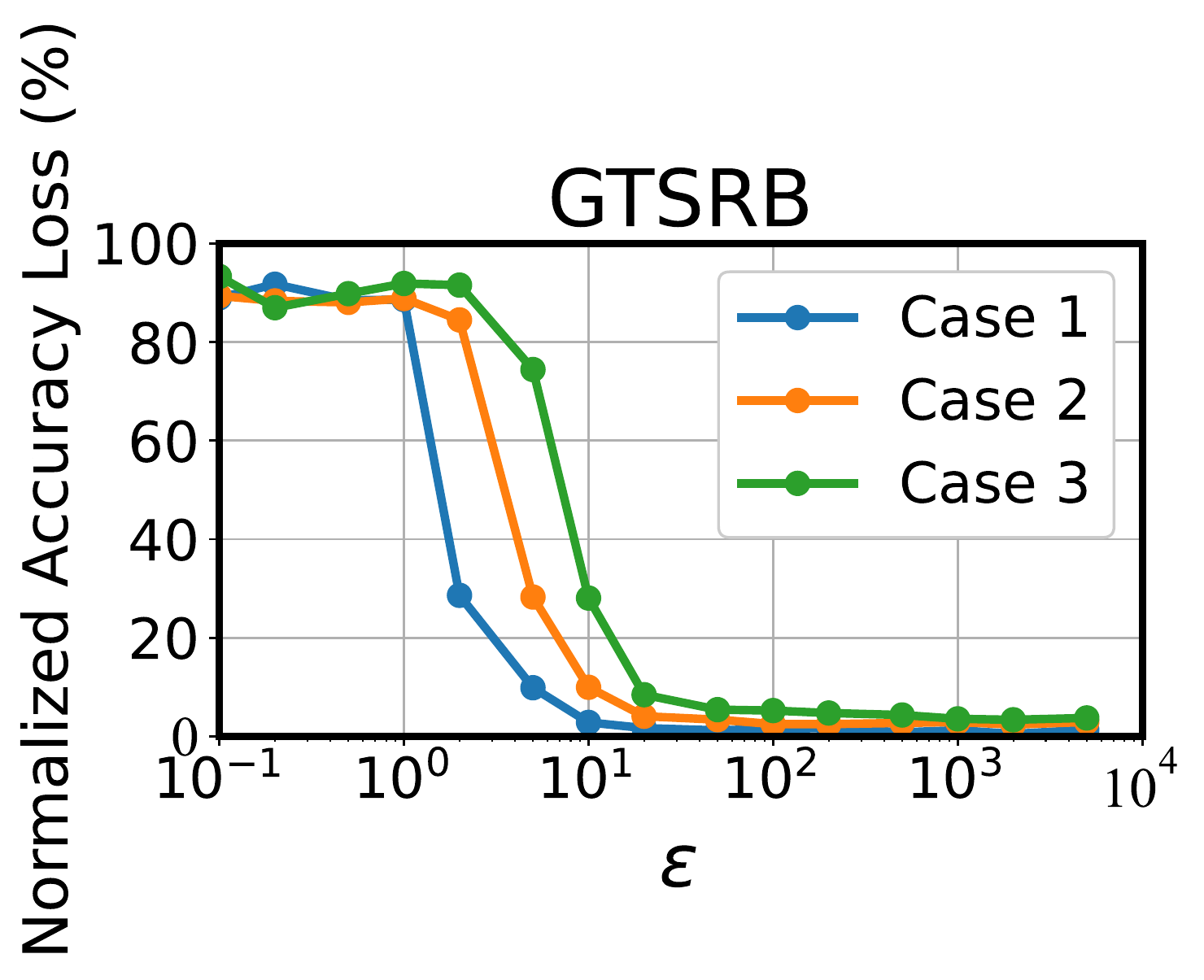}} \\ (b)
   \end{minipage}
\caption{Impact of the privacy budget $\epsilon$ on accuracy (GTSRB).}
\label{fig:gtsrb_acc}
\vspace{-15pt}
\end{figure}

\begin{figure*}[t!]
\centerline{\includegraphics[width=0.95\textwidth]{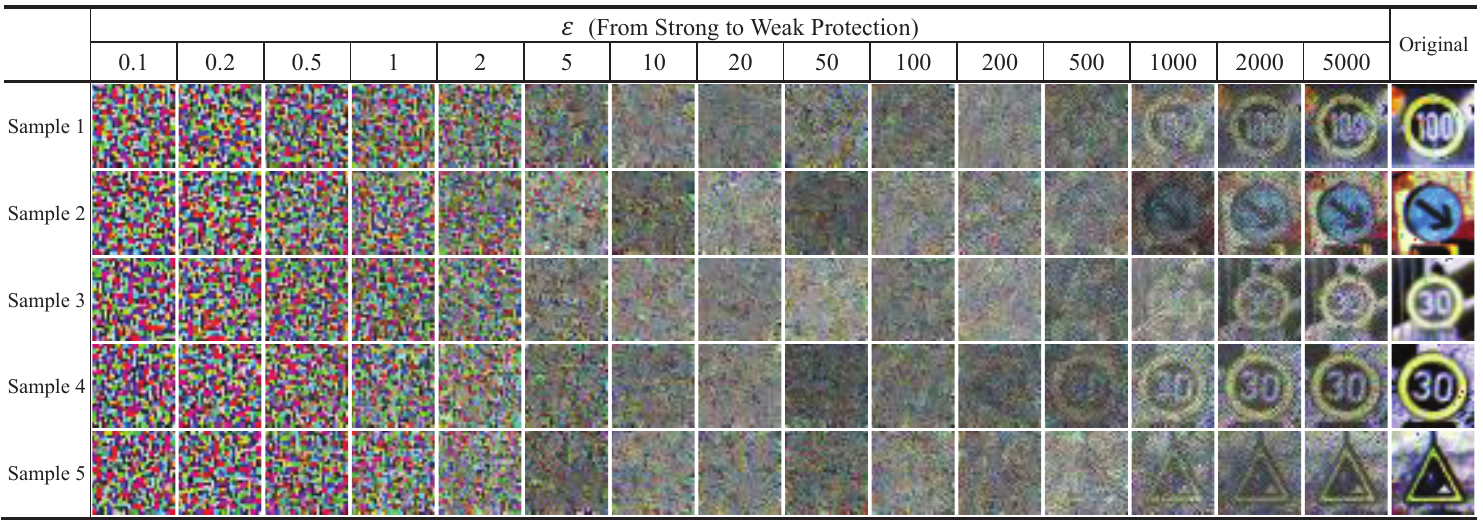}}
\caption{Visual results of applying the attack against the DP method (GTSRB Case 3).}
\label{fig:GTSRBattackedimage}
\end{figure*}

\begin{figure*}[t!]
\centerline{\includegraphics[width=1\textwidth]{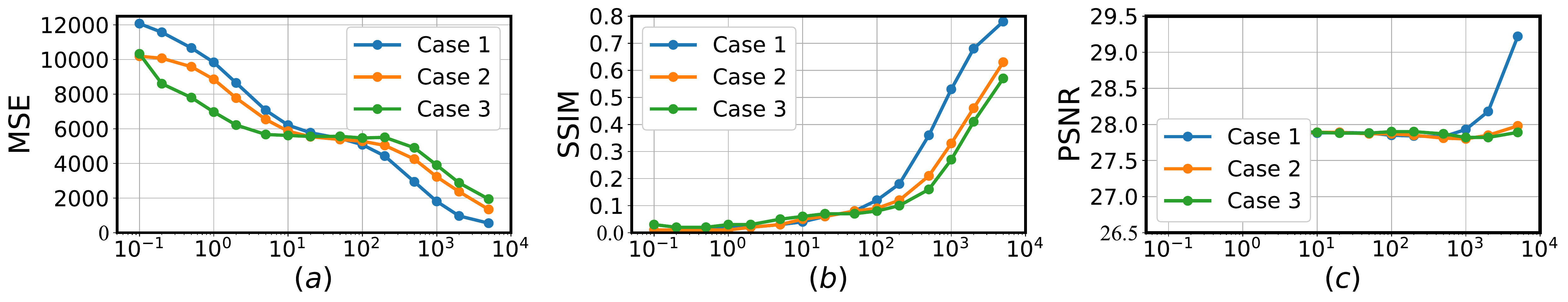}}
\caption{Evaluation of the quantitative metrics for the reconstruction attack efficacy (GTSRB): (a) MSE; (b) SSIM; (c) PSNR.}
\label{fig:quantified_results_gtsrb}
\end{figure*}

\paragraph{\bf Utility under DP} As shown in Table \ref{table:data_withoutaccuracy}, the baseline model over the SVHN dataset \emph{without} differential privacy (DP) achieves accuracy of $93.498\%$, $92.695\%$, $92.953\%$ in Cases 1, 2, and 3, respectively.
In Fig. \ref{fig:svhn_acc}, we show the accuracy results of the DP method (Fig. \ref{fig:svhn_acc} (a)) as well as the normalized accuracy loss (Fig. \ref{fig:svhn_acc} (b)) against the baseline accuracy, under varying values of the privacy budget $\epsilon$.
As depicted in the figure, the DNN model under the DP method has essentially no utility for $\epsilon <5$.
For $\epsilon \ge 5$, the accuracy achieved by the DP method is rapidly getting close to the baseline accuracy.
For instance, the accuracy results are $93.081\%$, $85.787\%$, $88.686\%$ for $\epsilon=5$ (normalized accuracy losses of $0.446\%$, $7.453\%$, $4.590\%$), $93.479\%$, $89.473\%$, $90.397\%$ for $\epsilon=10$ (normalized accuracy losses of $0.020\%$,  $3.476\%$,  $2.750\%$), $93.1928\%$, $91.8796\%$, $92.083\%$ for $\epsilon=100$ (normalized accuracy losses of $0.326\%$,  $0.880\%$,  $0.936\%$), and $93.199\%$, $92.083\%$, $92.249\%$ for $\epsilon=1000$ (normalized accuracy losses of $0.320\%$,  $0.660\%$,  $0.757\%$) for Case 1, Case 2, Case 3, respectively.
These results show that on the SVHN dataset the DP method can still achieve good accuracy highly close to the baseline accuracy under suitable $\epsilon$ values.

\begin{table}[t!]
\caption{Baseline Accuracy without Differential Privacy}
\centering
\begin{tabular}{@{}ccccc@{}}
\toprule
Model  & SVHN & GTSRB & STL-10 & CIFAR-10 \\ \midrule 
Case 1 & $93.498\%$ & $92.676\%$ & $69.576\%$ & $82.932\%$ \\
Case 2 & $92.695\%$ & $95.284\%$ & $67.448\%$ & $77.795\%$ \\
Case 3 & $92.953\%$ & $92.869\%$ & $67.333\%$ & $84.500\%$ \\
Average & $93.049\%$ & $93.610\%$ & $68.119\%$ & $81.742\%$ \\
\bottomrule
\end{tabular}
\label{table:data_withoutaccuracy}
\end{table}

\paragraph{\bf Protection Efficacy} We then examine the capability of the DP method in defending against the reconstruction attack.
In Fig. \ref{fig:SVHNattackedimage}, we show from a visual perspective the protection levels of differential privacy against the data reconstruction attack in Case 3 for some example testing images of the SVHN dataset. The results for Case 1 and 2 are shown in Appendix \ref{sec:AppendotherResulsts}.
That is, we show the original images and the reconstructed images derived by applying the attack to intermediate outputs of the local model part, with regard to varying privacy budget $\epsilon$.
As expected, the protection becomes less effective as the $\epsilon$ value increases.
According to the visual results in the figure, no meaningful information can be observed from the reconstructed images for $\epsilon \le 200$, indicating the DP method well protects the inputs against the reconstruction attack.
For $\epsilon \ge 500$, the visual information of some images can be (clearly) observed from the reconstructed images, such as Sample 3 and Sample 4.

In Fig. \ref{fig:quantified_results_svhn}, we show the evaluation of the results of the quantitative metrics (averaged over $100$ randomly chosen testing images), including the MSE, SSIM, and PSNR, with regard to varying privacy budget $\epsilon$.
For the MSE metric, a clear descending trend is observed for $\epsilon<10$.
Then, the MSE values become relatively stable for $10\le \epsilon \le 200$.
For $\epsilon > 200$, the MSE values decreasingly evolve, indicating the reconstructed images due to the attack are getting closer to the original images.
For the SSIM metric, overall there is an ascending trend, and a sharp increase can be observed for $\epsilon \ge 500$.
Regarding the PSNR metric, we observe that the PSNR values remain almost stable regardless of the varying privacy budget $\epsilon$.
No clear ascending trends can be observed with the increase of the privacy budget $\epsilon$ (except when $\epsilon$ is greater than 1000).
\textit{This suggests that PSNR is not an appropriate metric for measuring the resistance of the DP method against the attack in this context.}

\paragraph{\bf Note} From the above accuracy results and privacy measurement results, it is shown that on the SVHN dataset, the DNN model with the DP method, under suitable choices of $\epsilon$ values (e.g., $5 \le \epsilon \le 200$), can achieve accuracy comparable to the baseline while providing resistance to the reconstruction attack.

\subsection{Results over the GTSRB Dataset}

\paragraph{\bf Utility under DP} The baseline model over the GTSRB dataset \emph{without} differential privacy (DP) achieves accuracy of $92.676\%$, $95.284\%$, $92.869\%$ in Case 1, 2, and 3.
Fig. \ref{fig:gtsrb_acc} shows the accuracy results of the DP method (Fig. \ref{fig:gtsrb_acc} (a)) as well as the normalized accuracy loss (Fig. \ref{fig:gtsrb_acc} (b)) with respect to the baseline accuracy, under varying privacy budget $\epsilon$.
As depicted in the figure, the DNN model under the DP method has essentially no utility until $\epsilon$ exceeds $10$
For $\epsilon \ge 10$, the accuracy achieved by the DP method is becoming close to the baseline accuracy.
%
For instance, the accuracy results are $90.067 \%$, $85.811 \%$, $66.816 \%$ for $\epsilon=10$ (normalized accuracy losses of $2.815\%$,  $9.941\%$,  $28.053\%$), $91.287 \%$, $92.926 \%$, $ 88.025 \%$ for $\epsilon=100$ (normalized accuracy losses of $1.499\%$,  $2.475\%$,  $5.216\%$), and $91.533 \%$, $92.535 \%$, $89.587 \%$ for $\epsilon=1000$ (normalized accuracy losses of $1.233\%$, $2.885\%$,   $3.533\%$) in Case 1, Case 2, and Case 3, respectively.
These results show that on the GTSRB dataset the accuracy loss due to the DP method is small under suitable $\epsilon$ values.

\paragraph{\bf Protection Efficacy} Fig. \ref{fig:GTSRBattackedimage} shows from a visual perspective the protection levels of the DP method against the data reconstruction attack in Case 3 for some example testing images of the GTSRB dataset. Case 1 and 2 are shown in Appendix \ref{sec:AppendotherResulsts}.
As expected, the protection becomes less effective with the increase of the $\epsilon$ value.
According to the visual results in the figure, no meaningful information can be observed from the reconstructed images for $\epsilon \le 200$, indicating the DP method well protects the inputs against the reconstruction attack.
For $\epsilon \ge 500$, the visual information of the sample images can be (clearly) observed from the reconstructed images.

In Fig. \ref{fig:quantified_results_gtsrb}, we show the evaluation of the results of the quantitative metrics (averaged over $100$ randomly chosen testing images), including the MSE, SSIM, and PSNR, with regard to varying privacy budget $\epsilon$.
For the MSE metric, it reveals a clear descending trend for $\epsilon<10$.
Then, the MSE values become relatively stable for $10\le \epsilon \le 200$.
For $\epsilon > 200$, there is an obvious decrease in the MSE values, indicating the reconstructed images due to the attack are getting closer to the original images.
For the SSIM metric, there is an overall ascending trend, and a dramatic increase is shown for $\epsilon \ge 500$.
For the PSNR metric, we observe again that the PSNR values remain almost stable regardless of the privacy budget $\epsilon$.

\paragraph{\bf Note} From the above accuracy results and privacy measurement results, it is shown that over the GTSRB dataset, the DNN model with the DP method, under suitable choices of $\epsilon$ values (e.g., $100 \le \epsilon \le 200$), can achieve accuracy comparable to the baseline while providing resistance to the reconstruction attack.

\subsection{Results over the STL-10 Dataset}

\begin{figure}[!t]
\centering
   \begin{minipage}[t]{0.48\linewidth}
      \centering{\includegraphics[width=\linewidth]{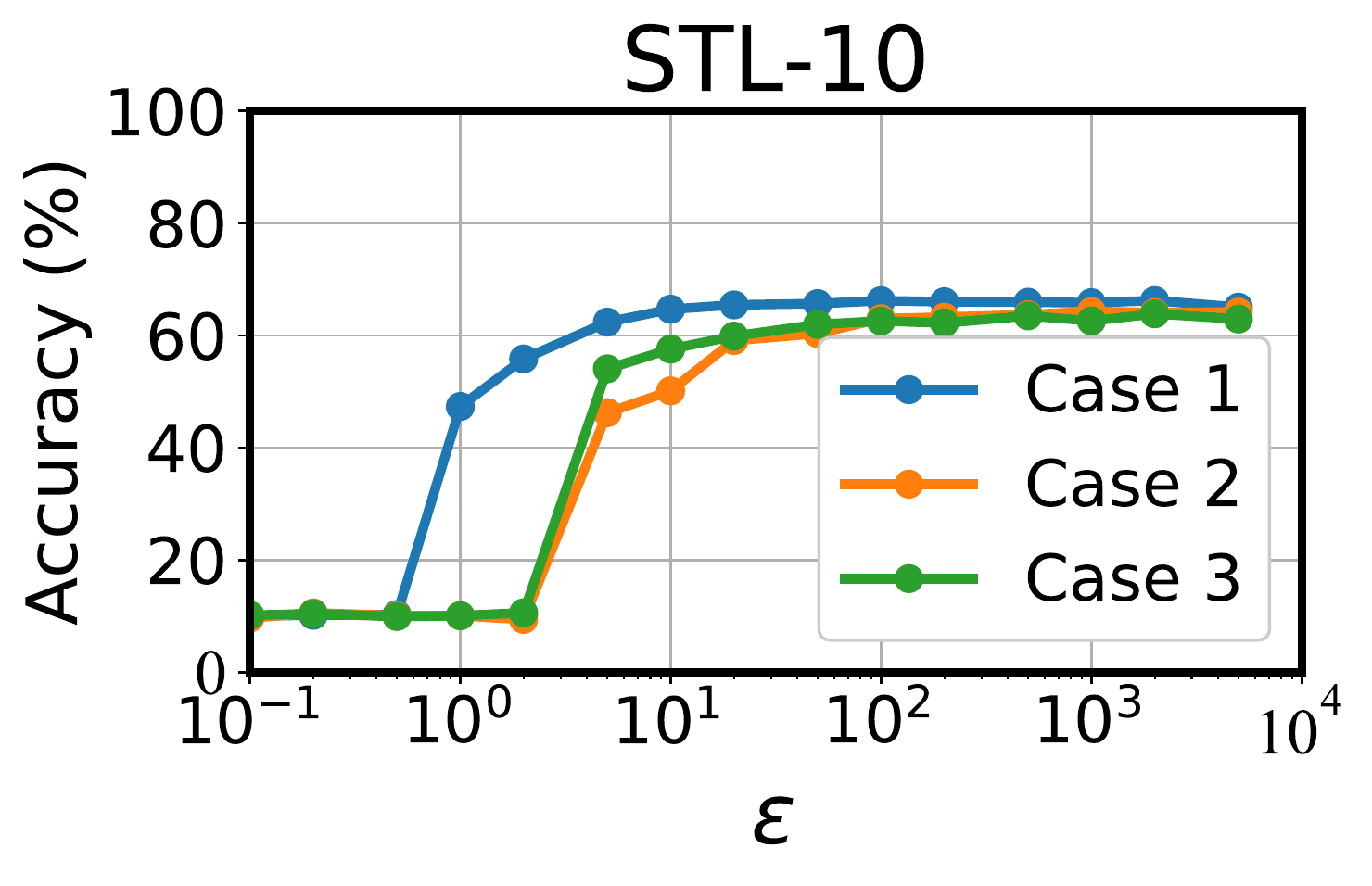}}\\ (a)
   \end{minipage}%
   \begin{minipage}[t]{0.47\linewidth}
      \centering{\includegraphics[width=\linewidth]{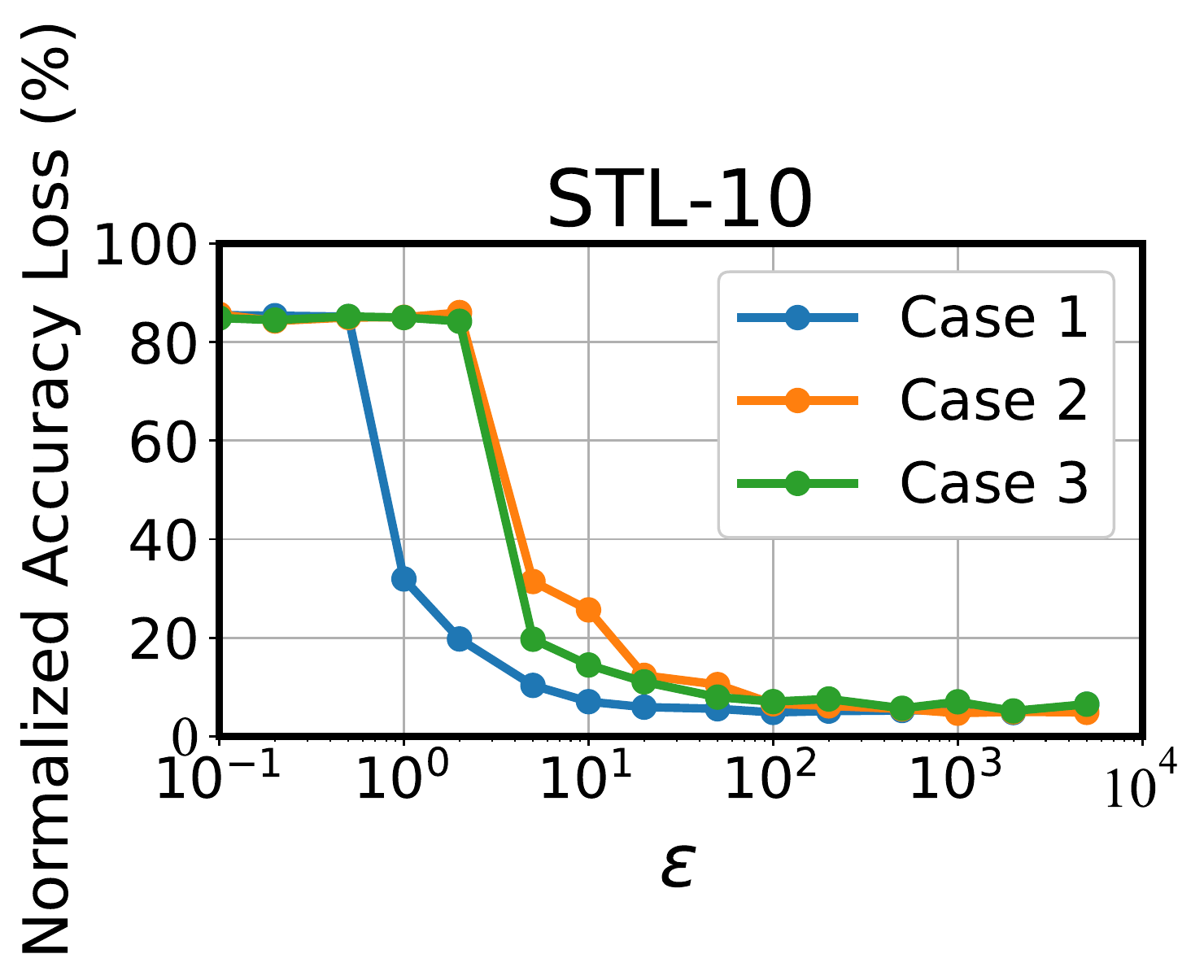}} \\ (b)
   \end{minipage}
\caption{Impact of the privacy budget $\epsilon$ on accuracy (STL-10).}
\label{fig:stl10_acc}
\vspace{-15pt}
\end{figure}

\begin{figure*}[t!]
\centerline{\includegraphics[width=0.95\textwidth]{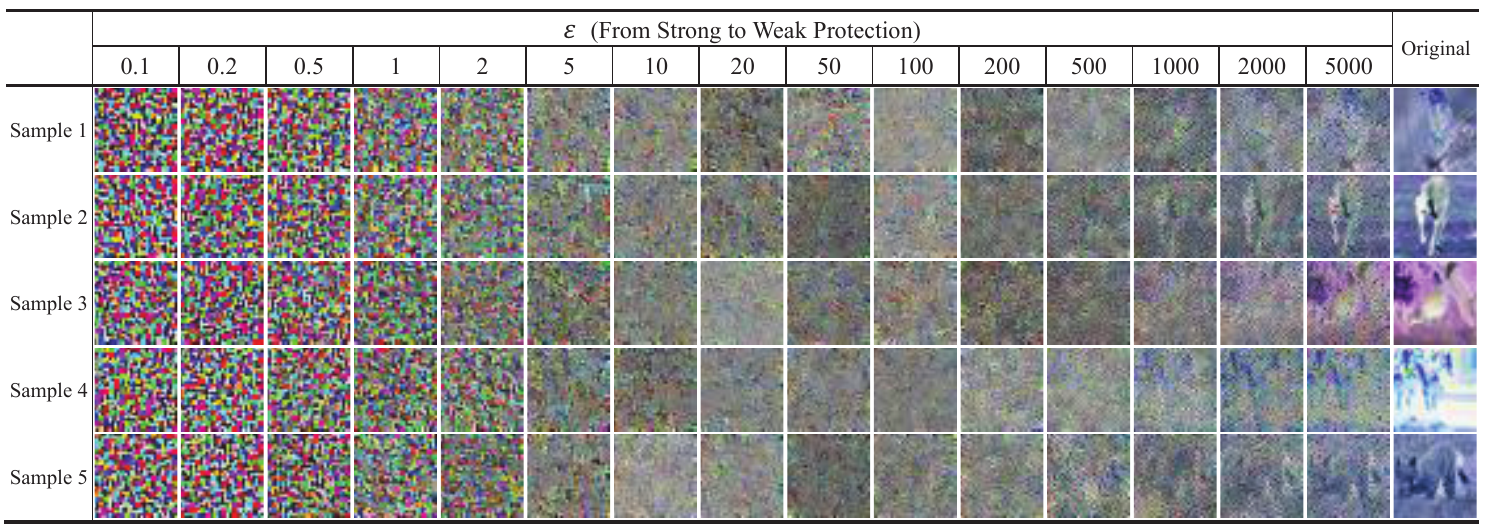}}
\caption{Visual results of applying the attack against the DP method (STL-10 Case 3).}
\label{fig:STL10attackedimage}
\end{figure*}

\begin{figure*}[t!]
\centerline{\includegraphics[width=1\textwidth]{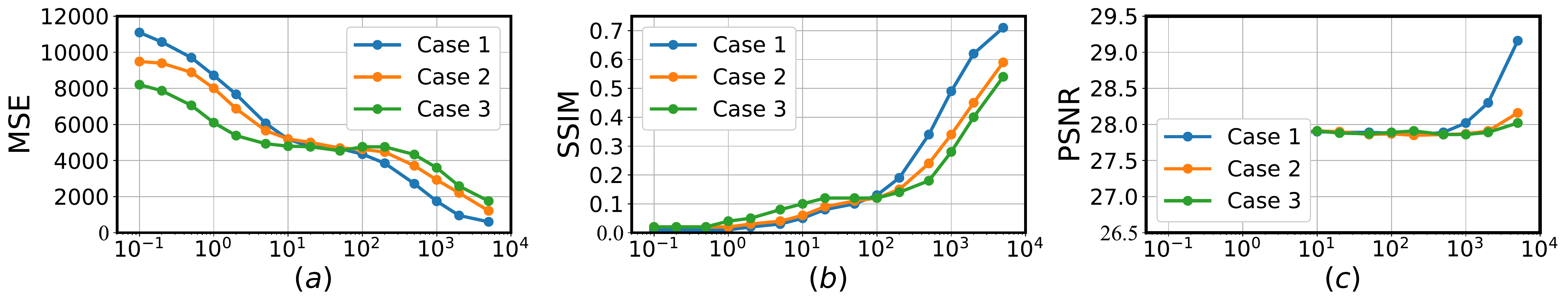}}
\caption{Evaluation of the quantitative metrics for the reconstruction attack efficacy (STL-10): (a) MSE; (b) SSIM; (c) PSNR.}
\label{fig:quantified_results_stl10}
\end{figure*}

\paragraph{\bf Utility under DP} The baseline model over the STL-10 dataset \emph{without} differential privacy (DP) achieves accuracy of $69.576\%$, $67.448\%$, $67.333\%$ in Case 1, 2, and 3.
Such accuracy levels also appeared in prior work \cite{dosovitskiy2014discriminative}, and is orthogonal to our study in this paper.
In Fig. \ref{fig:stl10_acc}, we show the accuracy results of the DP method (Fig. \ref{fig:stl10_acc} (a)) as well as the normalized accuracy loss (Fig. \ref{fig:stl10_acc} (b)) against the baseline accuracy, under varying values of the privacy budget $\epsilon$.
As shown, the DNN model under the DP method has essentially no utility for $\epsilon <10$.
For $\epsilon \ge 10$, the accuracy achieved by the DP method is getting close to the baseline accuracy.
For instance, the accuracy results are $64.690 \%$, $50.149 \%$, $57.593 \%$ for $\epsilon=10$ (normalized accuracy losses of $7.023\%$,  $25.648\%$, $14.465\%$), $66.207 \%$, $63.003 \%$, $62.598 \%$ for $\epsilon=100$ (normalized accuracy losses of $4.842\%$,   $6.590\%$,  $7.032\%$), and $65.857 \%$, $64.289 \%$, $62.621 \%$ for $\epsilon=1000$ (normalized accuracy losses of $5.345\%$,  $4.684\%$,  $6.998\%$) in Case 1, Case 2, and Case 3, respectively.
These results show that on the STL-10 dataset, the DP method can achieve accuracy comparable to the baseline under suitable $\epsilon$ values.

\paragraph{\bf Protection Efficacy} Fig. \ref{fig:STL10attackedimage} shows from a visual perspective the protection levels of the DP method against the data reconstruction attack in Case 3 for some example testing images of the STL-10 dataset. Case 1 and 2 are shown in Appendix \ref{sec:AppendotherResulsts}.
As expected, the protection becomes less effective with the increase of the $\epsilon$ value.
It is observed that even at $\epsilon=1000$, the reconstructed images almost reveal no meaningful visual information of the original images.
In Fig. \ref{fig:quantified_results_stl10}, we show the evaluation of the results of the quantitative metrics (averaged over $100$ randomly chosen testing images), including the MSE, SSIM, and PSNR, with regard to varying privacy budget $\epsilon$.
For the MSE metric, a clear descending trend is observed for $\epsilon<10$.
Then, the MSE values become relatively stable for $10\le \epsilon \le 200$.
For $\epsilon > 200$, the MSE values decreasingly evolve, indicating the reconstructed images due to the attack are getting closer to the original images.
For the SSIM metric, overall there is an ascending trend, and a sharp increase can be observed for $\epsilon \ge 500$.
Regarding the PSNR metric, it is shown again that the PSNR values remain almost stable regardless of the varying privacy budget $\epsilon$.

\paragraph{\bf Note} From the above accuracy results and privacy measurement results, it is shown that over the STL-10 dataset, the DNN model with the DP method, under suitable choices of $\epsilon$ values (e.g., $100 \le \epsilon \le 500$), can achieve accuracy comparable to the baseline while protecting the input privacy.

\subsection{Results over the CIFAR-10 Dataset}

\begin{figure}[!t]
\centering
   \begin{minipage}[t]{0.48\linewidth}
      \centering{\includegraphics[width=\linewidth]{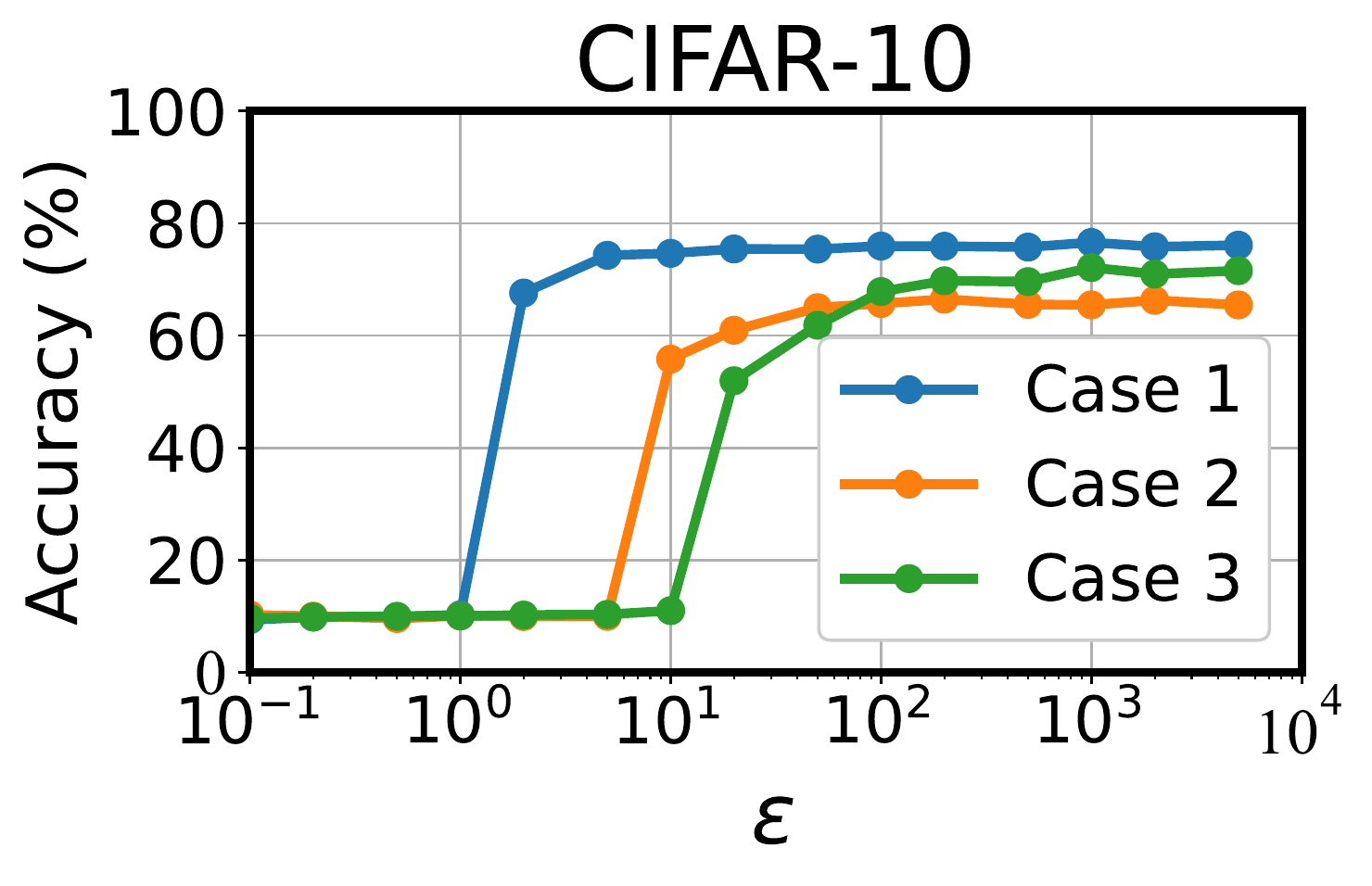}}\\ (a)
   \end{minipage}%
   \begin{minipage}[t]{0.47\linewidth}
      \centering{\includegraphics[width=\linewidth]{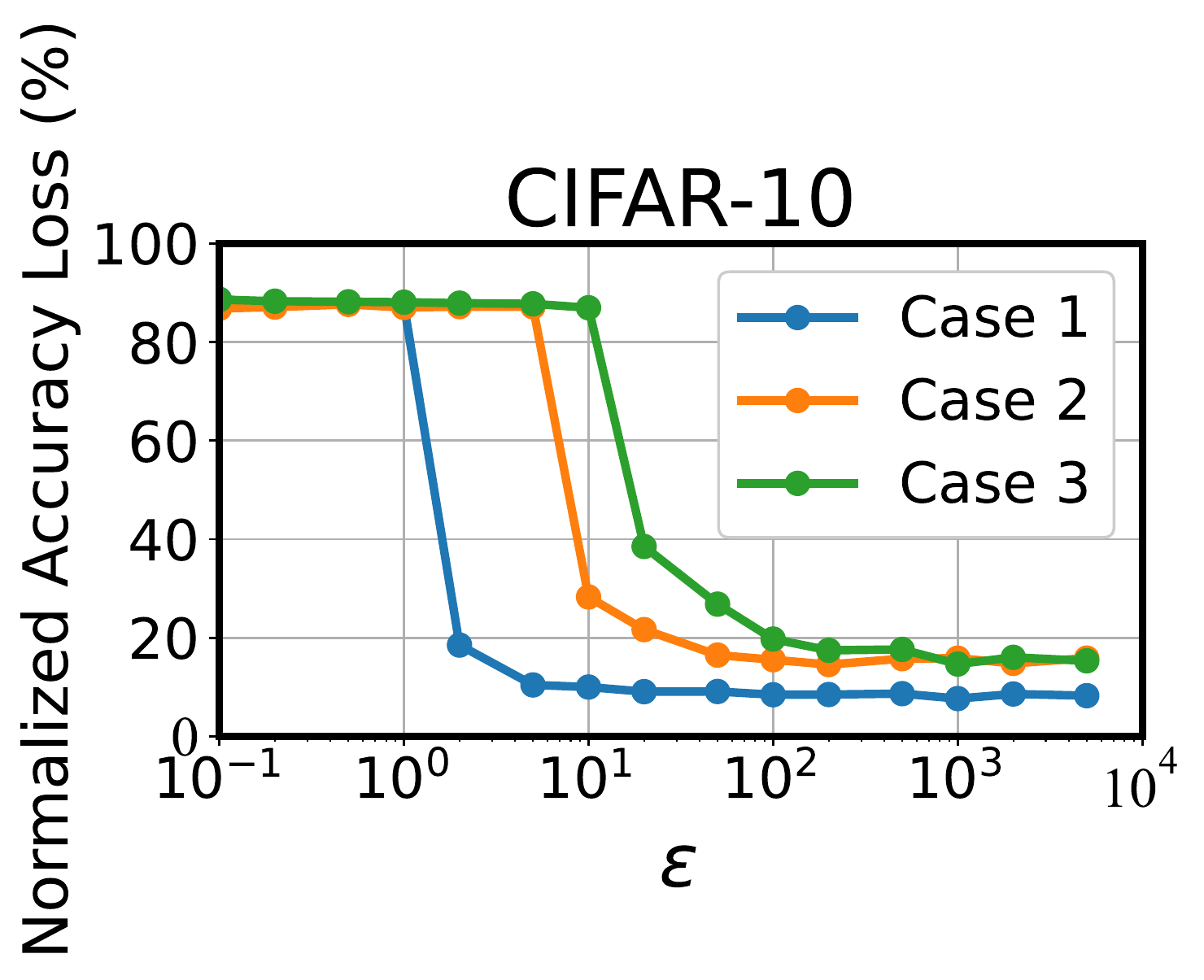}} \\ (b)
   \end{minipage}
\caption{Impact of the privacy budget $\epsilon$ on accuracy (CIFAR-10).}
\label{fig:cifar10_acc}
\vspace{-15pt}
\end{figure}

\begin{figure*}[t!]
\centerline{\includegraphics[width=0.95\textwidth]{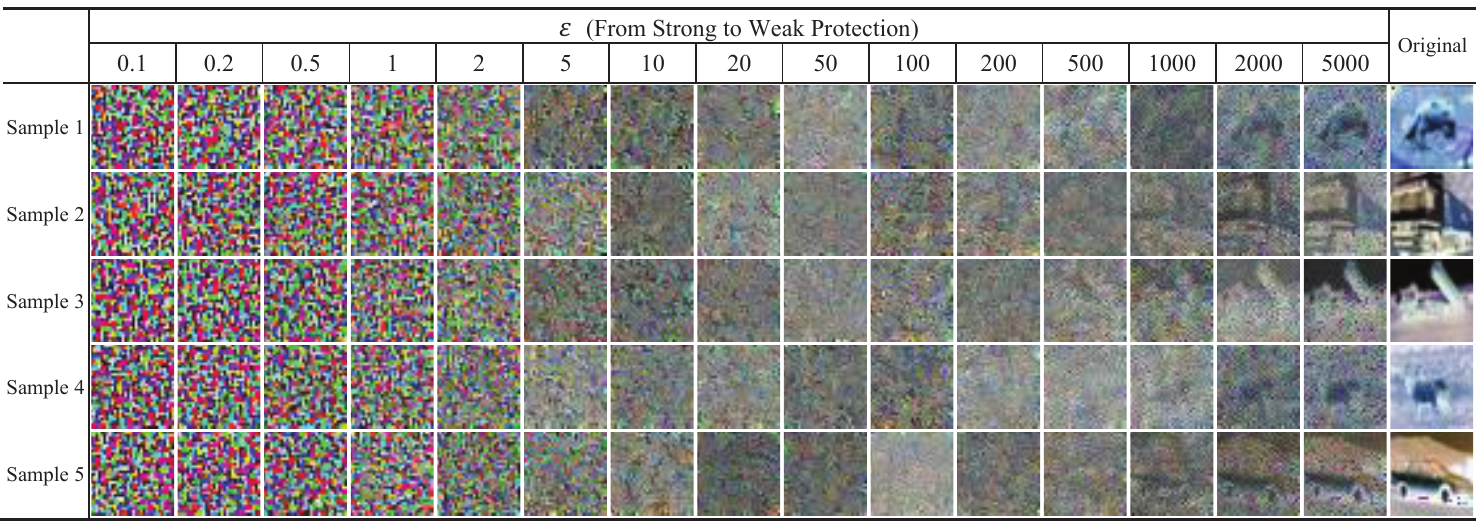}}
\caption{Visual results of applying the attack against the DP method (CIFAR-10 Case 3).}
\label{fig:CIFAR10attackedimage}
\end{figure*}

\begin{figure*}[t!]
\centerline{\includegraphics[width=1\textwidth]{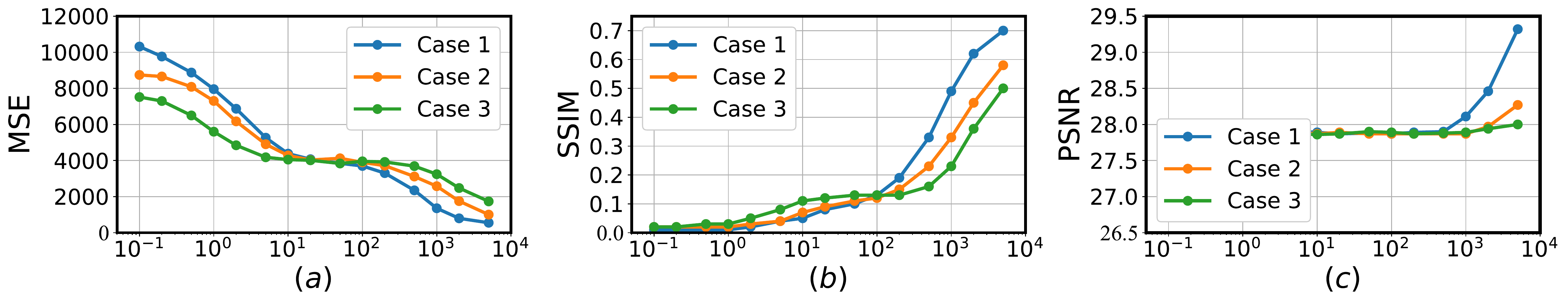}}
\caption{Evaluation of the quantitative metrics for the reconstruction attack efficacy (CIFAR-10): (a) MSE; (b) SSIM; (c) PSNR.} 
\label{fig:quantified_results_cifar10}
\end{figure*}

\paragraph{\bf Utility under DP} The baseline model over the CIFAR-10 dataset \emph{without} differential privacy (DP) achieves accuracy of $82.932\%$, $77.795\%$, $84.5\%$ in Case 1, 2, and 3.
We show in Fig. \ref{fig:cifar10_acc} the accuracy results of the DP method (Fig. \ref{fig:cifar10_acc} (a)) as well as the normalized accuracy loss (Fig. \ref{fig:cifar10_acc} (b)) against the baseline accuracy, with regard to varying privacy budget $\epsilon$.
As shown in the figure, the DNN model under the DP method has almost no utility for $\epsilon <50$.
For $\epsilon \ge 200$, the accuracy does not increase significantly.
In particular, for $200 \le\epsilon \le 1000$, the accuracy varies from $75.905\%$, $66.475\%$, $69.755\%$ (normalized accuracy losses of $8.473\%$,  $14.551\%$, $17.450\%$) to $76.588\%$, $65.462\%$, $72.114\%$ (normalized accuracy losses of $7.650\%$, $15.853\%$,  $14.658\%$), which is not close to the baseline accuracy of $82.932\%$, $77.795\%$, $84.5\%$ in Case 1, Case 2, and Case 3, respectively.
These results show that on the CIFAR-10 dataset, the DP method can retain meaningful utility of the DNN model yet the accuracy loss against the base accuracy is notable.

%
\paragraph{\bf Protection Efficacy} Fig. \ref{fig:CIFAR10attackedimage} shows from a visual perspective the protection levels of the DP method against the data reconstruction attack in Case 3 for some example testing images of the CIFAR-10 dataset. Case 1 and 2 are shown in Appendix \ref{sec:AppendotherResulsts}.

According to the visual results in the figure, no meaningful information can be observed from the reconstructed images for $\epsilon \le 500$, indicating the DP method well protects the inputs against the reconstruction attack.
Fig. \ref{fig:quantified_results_cifar10} shows the evaluation of the results of the quantitative metrics (averaged over $100$ randomly chosen testing images), including the MSE, SSIM, and PSNR, with regard to varying privacy budget $\epsilon$.
For the MSE metric, a clear descending trend is observed for $\epsilon<10$.
Then, the MSE values become relatively stable for $10\le \epsilon \le 200$.
For $\epsilon > 200$, the MSE values decreasingly evolve, indicating the reconstructed images due to the attack are getting closer to the original images.
For the SSIM metric, overall there is an ascending trend, and a sharp increase can be observed for $\epsilon \ge 500$.
Regarding the PSNR metric, we observe that the PSNR values remain almost stable regardless of the varying $\epsilon$.

\paragraph{\bf Note} From the above accuracy results and privacy measurement results, it is shown that over the CIFAR-10 dataset, the DNN model with the DP method, under suitable choices of $\epsilon$ values (e.g., $200 \le \epsilon \le 500$), can only retain a meaningful utility of the DNN model while providing resistance to the reconstruction attack.

\begin{figure}[!t]
\centerline{\includegraphics[width=0.4\textwidth]{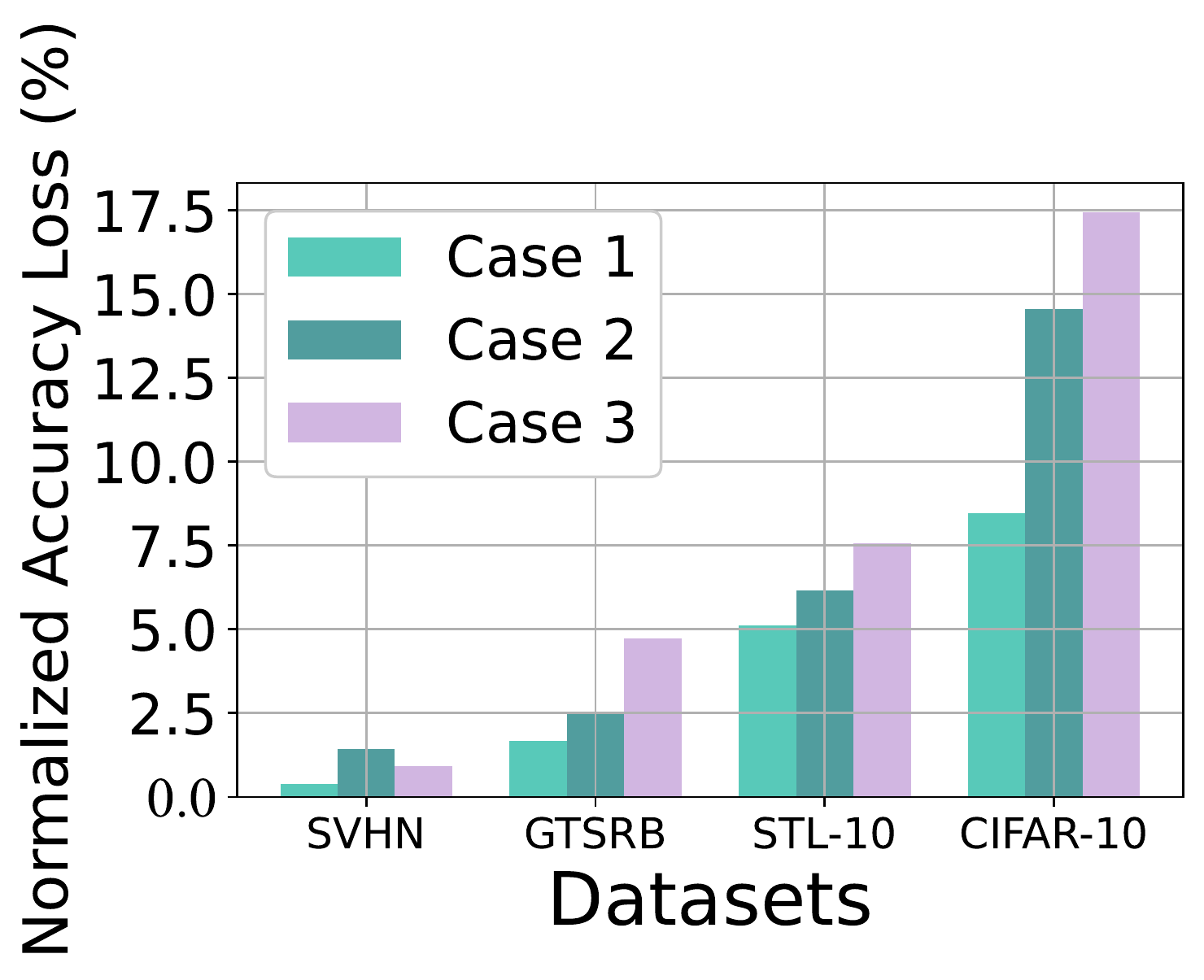}}
\caption{Comparison of normalized accuracy drops over different datasets. ($\epsilon$ = 200)} 
\label{fig:acc_loss_comparison200}
\vspace{-12pt}
\end{figure}


\section{Insights and Discussions}
\label{sec:discussion}

In response to our research question above on whether the differential privacy framework is able to protect collaborative inference while preserving utility, we discuss our findings and draw insights as follows.

\noindent \textbf{Differential privacy is usable for collaborative inference in the presence of the data reconstruction attack.}
From our results above, we consistently observe that the use of differential privacy can retain the (meaningful) usability of the DNN model, while providing protection on the input privacy in collaborative inference.
For different datasets, however, our observation is that the suitable intervals of the privacy budget $\epsilon$ that can protect the input privacy while maintaining good accuracy could vary.
For example, on the SVHN dataset, for $\epsilon=5$, the (normalized) accuracy loss is $0.446\%$, $7.453\%$, $4.590\%$ in Case 1, 2, and 3 while it is $9.854\%$, $28.257\%$, $74.435\%$ in Case 1, 2, and 3 on the GTSRB dataset.
On the GTSRB dataset, for $\epsilon=500$, the visual information of original images can be observed from the reconstructed images, while no meaningful visual information from the attack can be observed on the CIFAR-10 dataset.
Overall, across all the datasets being evaluated, our empirical observation is that the interval $100\le \epsilon \le 200$ tends to provide a good trade-off between utility and privacy protection.

\noindent \textbf{Whether differential privacy can achieve accuracy close to the baseline is dataset-dependent.}
From the results over the four datasets, we observe that on the SVHN, GTSRB, and STL-10 datasets, the use of differential privacy is able to achieve accuracy close to the non-private baseline. 
For example, as shown in Fig. \ref{fig:acc_loss_comparison200}, for $\epsilon=200$ where privacy protection is ensured, the (normalized) accuracy loss is $0.395\%$, $1.430\%$, $0.928\%$ on SVHN, $1.671\%$, $2.464\%$, $4.720\%$ on GTSRB, and $5.106\%$, $6.161\%$, $7.567\%$ on STL-10 respectively, while it is up to $8.473\%$, $14.551\%$, $17.450\%$ on CIFAR-10, for Case 1, Case 2, and Case 3, respectively.

On CIFAR-10, even when $\epsilon$ further increases to $2000$ or $5000$ where input privacy is compromised as shown in Fig. \ref{fig:CIFAR10attackedimage}, the accuracy loss still stays at a high level, i.e., $8.581\%$, $14.794\%$, $16.024\%$ for $\epsilon=2000$, and $8.241\%$, $15.868\%$, $15.315\%$ for $\epsilon=5000$.
Hence, we point out that even differential privacy can retain the (meaningful) usability of the DNN model in collaborative inference, it may not always be able to maintain the accuracy comparable to the non-private baseline.

\paragraph{Empirical Guide}  Our empirical insight is that differential privacy appears to perform better for datasets with small intra-class variation in collaborative inference, since according to our observation CIFAR-10 has relatively large intra-class variation compared to the other datasets. Specifically, it is visually observable that the order of intra-class variation of the four tested datasets is as follows: CIFAR10$>$STL-10$>$GTSRB$>$SVHN. Accordingly, the averaged accuracy drops across different splitting cases due to differential privacy are 12.454\%, 5.021\%, 2.066\%, 0.476\% for CIFAR-10, STL-10, GTSRB, and SVHN, respectively, given the largest tested privacy budget per each dataset that can still provide protection against the reconstruction attack ($\epsilon=200$ for SVHN and GTSRB, and $\epsilon=500$ for STL-10 and CIFAR-10, as visually observed).

One simple criterion for intra-class variation is that the more specific the class is, the smaller the intra-class variation will be. For instance, the intra-class variation of German Shepherd Dog class is smaller than the intra-class of dog class, since the latter is more general.
We hope our initial study can stimulate research activities for further in-depth investigation.

\paragraph{Potential Reason} When the intra-class variation becomes larger, the sensitivity to the noise injected from the differential privacy could be higher. This could lead to notable degradation in the accuracy. A formal proof and corroboration in this direction is an interesting future work.

\section{Related Work}
\label{sec:related_work}

The user privacy issues have been extensively studied \cite{huang2020privacy,nguyen2020privacy,andreoletti2020privacy,dong2021network,khan2021socially,zhang2016fakemask,subramanya2021centralized,ding2019extended,groleat2012distributed}.
FakeMask \cite{zhang2016fakemask} proposed a technology to protect users' privacy by disclosing fake contexts to solve the privacy problem on sensor-equipped smartphones. The work \cite{huang2020privacy} proposed a privacy protection scheme based on a differential privacy model combined with clustering and randomization algorithms.
In particular, there are privacy methods for machine learning models, such as \cite{nguyen2020privacy,andreoletti2020privacy,dong2021network,khan2021socially,subramanya2021centralized,groleat2012distributed}.
A reinforcement learning algorithm that guarantees privacy in the optimization of the Markov decision-making process and can efficiently solve a large state space in a blockchain scenario by proposing a reinforcement learning-based offloading method was developed in \cite{nguyen2020privacy}.
The optimization method of the Deep Reinforcement Learning algorithm for detecting abnormal traffic that can monitor network transmission in real-time using anomaly detection and effectively detects external attacks is suggested in \cite{dong2021network}.
In addition, the works \cite{khan2021socially,subramanya2021centralized,ZhengLLYYW22,ZhuLLYXL22} used federated learning for privacy protection in training models over distributed datasets.
%
%
There is also a line of work \cite{ZhengDTWZ21,LiuZYY21} on leveraging cryptographic techniques to secure DNN inference.

Our work is related to prior works on evaluating the effectiveness of differential privacy in machine learning with attacks.
In \cite{RahmanRLM18}, Rahman \textit{et al.} evaluate membership inference attacks against a differentially private DNN model which is proposed in \cite{AbadiCGMMT016}.
In \cite{Jayaraman019}, Jayaraman and Evans study the effectiveness of different relaxed notions of differential privacy which are proposed for training differentially private machine learning models, against membership inference attacks and attribute inference attacks.
In \cite{abs-1912-11328}, Bernau \textit{et al.} compare local and central differential privacy mechanisms under membership inference attacks.
All these works are proposed for the scenario where differential privacy is employed to protect the privacy of \emph{training data}.
Different from prior works, we present the first study on evaluating differential privacy when it is leveraged to protect the privacy of \emph{model inputs} in collaborative inference, against the state-of-the-art data reconstruction attack.

\section{Conclusion and Future Work}
\label{sec:conclusion}

In this paper, we initiate the first comprehensive study on the assessment of the practical usability of differential privacy for collaborative inference in the presence of state-of-the-art data reconstruction attack.
We conduct an extensive empirical evaluation over four datasets, examining the impact of varying privacy budget $\epsilon$ on the aspects including inference accuracy, visual protection strengths, and quantitative metrics.
Our results reveal that differential privacy can be usable in the presence of the reconstruction attack under certain conditions.
Practical insights and guidelines on the privacy-utility trade-offs have been drawn when deploying differential privacy for collaborative inference in practice. More specifically, an easy-to-adopt drawn guideline is that smaller intra-class variation of the dataset, more pragmatic of the DP for collaborative inference.
We hope our work can lead to a deeper understanding of the effectiveness of using differential privacy for the protection of model input privacy in collaborative inference for IoT applications.

For furture work, it is interesting to explore quantitative measures for capturing dataset characteristics (e.g., intra-class variation) so as to better study the relation between dataset characteristics and the protection strengths of differntial privacy.
It is also interesting to extend our study to non-image data, if reconstruction attacks against non-image data emerge in future.

\section*{Acknowledgements}
This paper was supported in part by the Guangdong Basic and Applied Basic Research Foundation under Grant 2021A1515110027, in part by the Shenzhen Science and Technology Program under Grant RCBS20210609103056041, in part by the National Natural Science Foundation of China under Grant 62002167, in part by the Natural Science Foundation of JiangSu under Grant BK20200461, in part by the Research Grants Council of Hong Kong under Grants CityU 11217819, 11217620, RFS2122-1S04, N\_CityU139/21, C2004-21GF, R1012-21, and R6021-20F, in part by the Shenzhen Municipality Science and Technology Innovation Commission under Grant SGDX20201103093004019, and in part by the Information \& Communications Technology Promotion Grant funded by the Korea government with Grant No. 2022-0-01199.

\section*{Declarations}
\begin{itemize}
\item \textbf{Funding} This paper was supported in part by the Guangdong Basic and Applied Basic Research Foundation under Grant 2021A1515110027, in part by the Shenzhen Science and Technology Program under Grant RCBS20210609103056041, in part by the National Natural Science Foundation of China under Grant 62002167, in part by the Natural Science Foundation of JiangSu under Grant BK20200461, in part by the Research Grants Council of Hong Kong under Grants CityU 11217819, 11217620, RFS2122-1S04, N\_CityU139/21, C2004-21GF, R1012-21, and R6021-20F, in part by the Shenzhen Municipality Science and Technology Innovation Commission under Grant SGDX20201103093004019, and in part by the Information \& Communications Technology Promotion Grant funded by the Korea government with Grant No. 2022-0-01199.
\item \textbf{Competing interests} The authors have no relevant financial or non-financial interests to disclose. 
\item \textbf{Ethics approval} This article does not contain any studies with human participants performed by any of the authors.

\item \textbf{Data availability} The datasets used during this study are publicly available, and the references to their sources have been given in this published article.
\item \textbf{Authors' contributions} Conceptualization: Jihyeon Ryu, Yifeng Zheng, Yansong Gao, Alsharif Abuadbba; Methodology: Jihyeon Ryu, Yifeng Zheng, Yansong Gao, Alsharif Abuadbba; Formal analysis and investigation: Jihyeon Ryu, Yifeng Zheng, Yansong Gao; Writing - original draft preparation: Jihyeon Ryu, Yifeng Zheng, Yansong Gao, Alsharif Abuadbba; Writing - review and editing: Junyaup Kim, Dongho Won, Surya Nepal, Hyoungshick Kim, Cong Wang; Funding acquisition: Yifeng Zheng, Yansong Gao.
\end{itemize}

\bibliographystyle{IEEEtran}
\bibliography{references}

\appendix

\begin{figure*}[!h]
\centering
   \vspace{15pt}
   \begin{minipage}[t]{0.35\linewidth}
      \centering{\includegraphics[width=\linewidth]{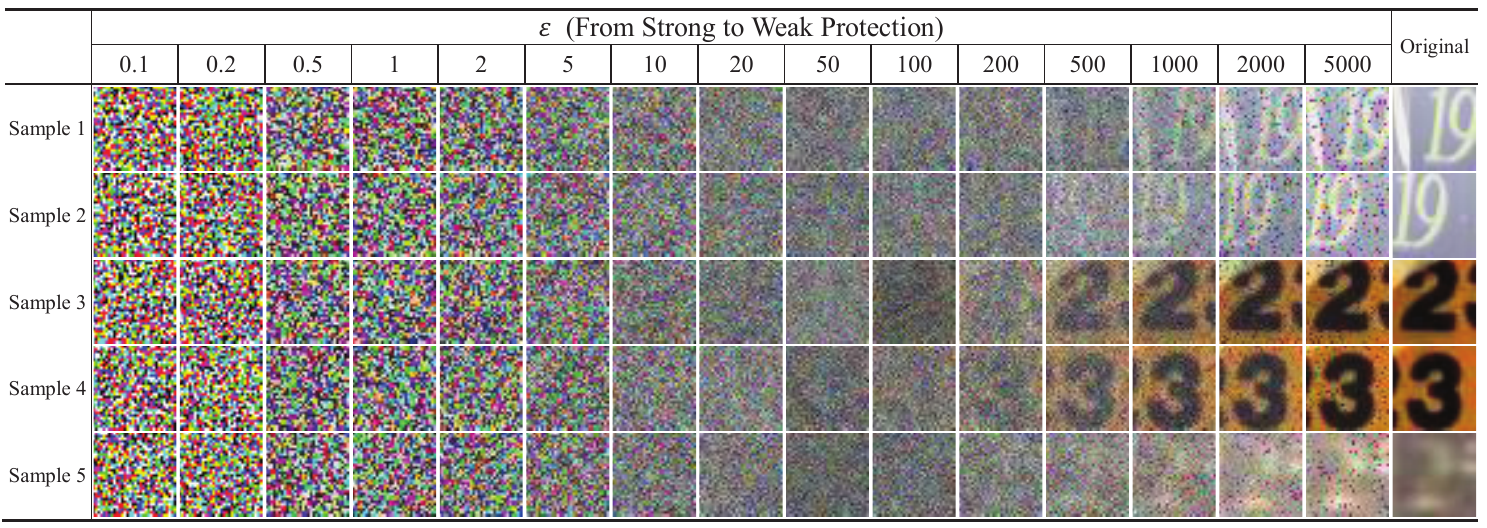}} \\ SVHN Case 1
   \end{minipage}%
   \vspace{10pt}
   \begin{minipage}[t]{0.35\linewidth}
      \centering{\includegraphics[width=\linewidth]{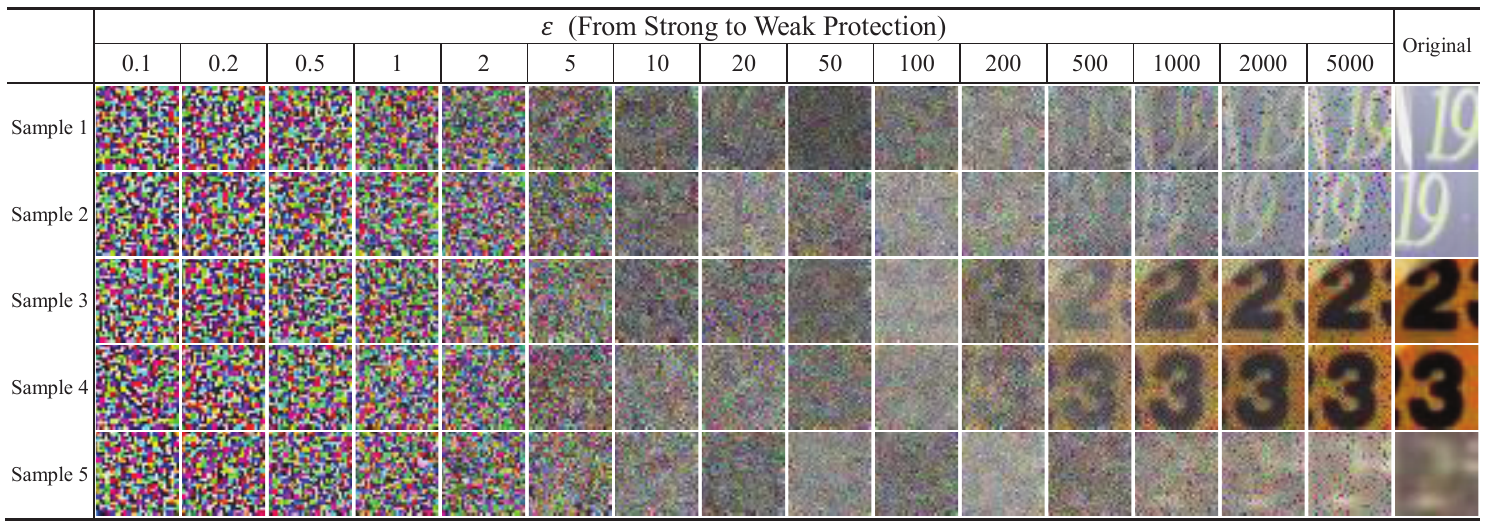}} \\ SVHN Case 2
   \end{minipage}
   \vspace{10pt}
   \begin{minipage}[t]{0.35\linewidth}
      \centering{\includegraphics[width=\linewidth]{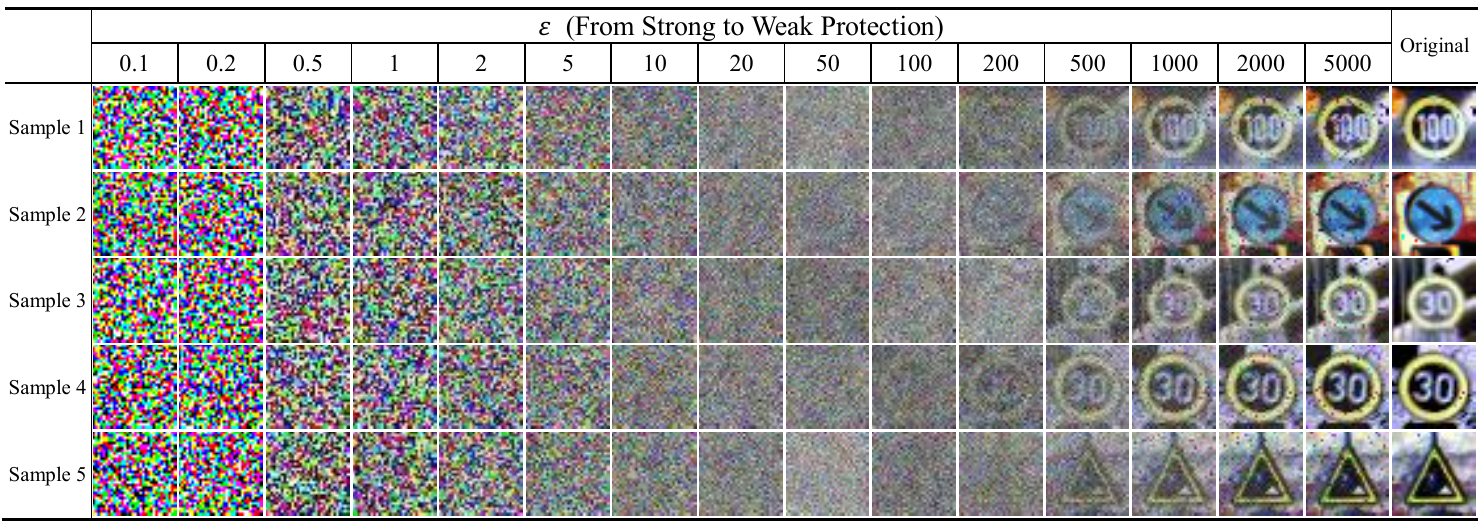}} \\ GTSRB Case 1
   \end{minipage}
   \begin{minipage}[t]{0.35\linewidth}
      \centering{\includegraphics[width=\linewidth]{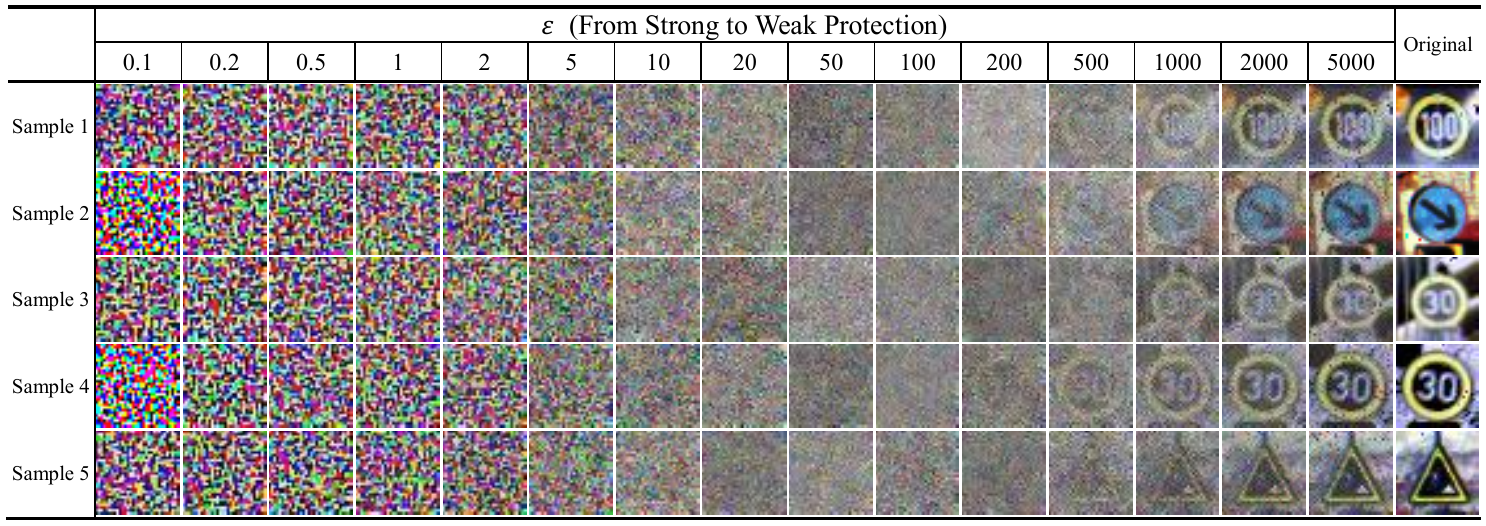}} \\ GTSRB Case 2
   \end{minipage}
   \vspace{10pt}
   \begin{minipage}[t]{0.35\linewidth}
      \centering{\includegraphics[width=\linewidth]{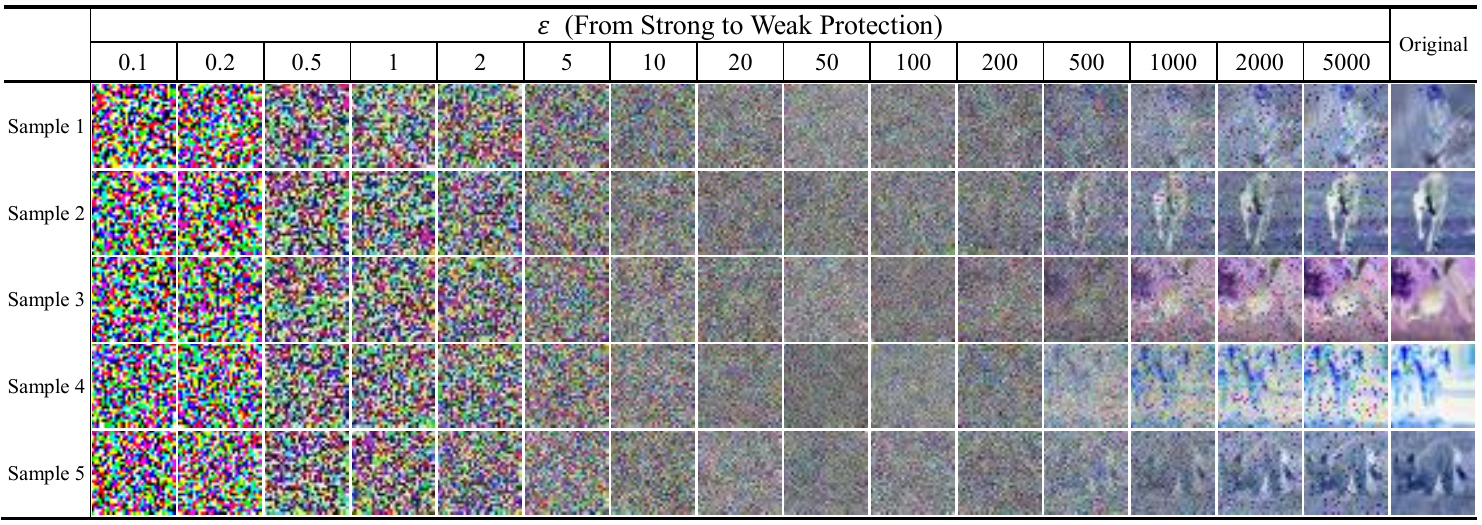}} \\ STL-10 Case 1
   \end{minipage}
   \begin{minipage}[t]{0.35\linewidth}
      \centering{\includegraphics[width=\linewidth]{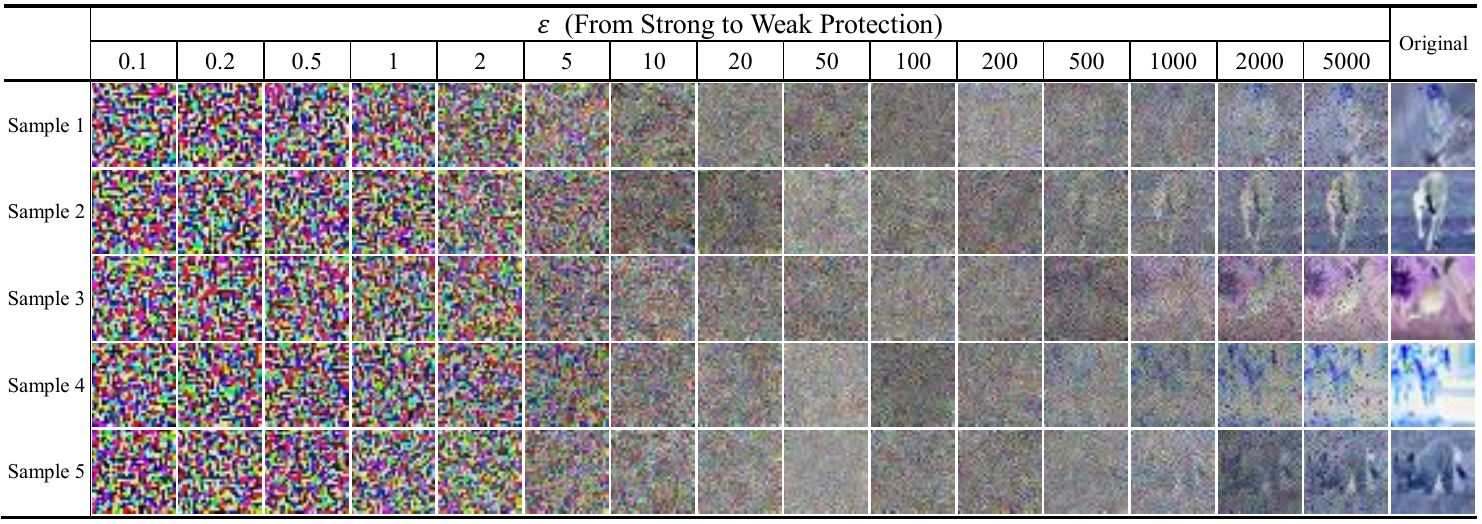}} \\ STL-10 Case 2
   \end{minipage}
   \vspace{10pt}
   \begin{minipage}[t]{0.35\linewidth}
      \centering{\includegraphics[width=\linewidth]{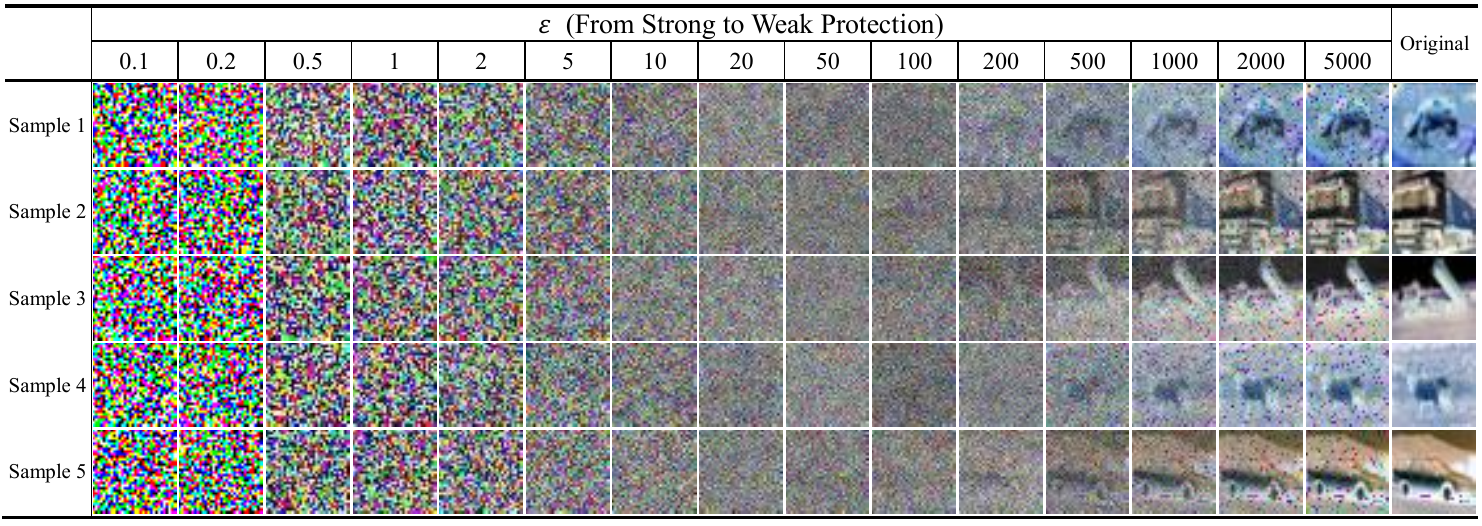}} \\ CIFAR-10 Case 1
   \end{minipage}
   \begin{minipage}[t]{0.35\linewidth}
      \centering{\includegraphics[width=\linewidth]{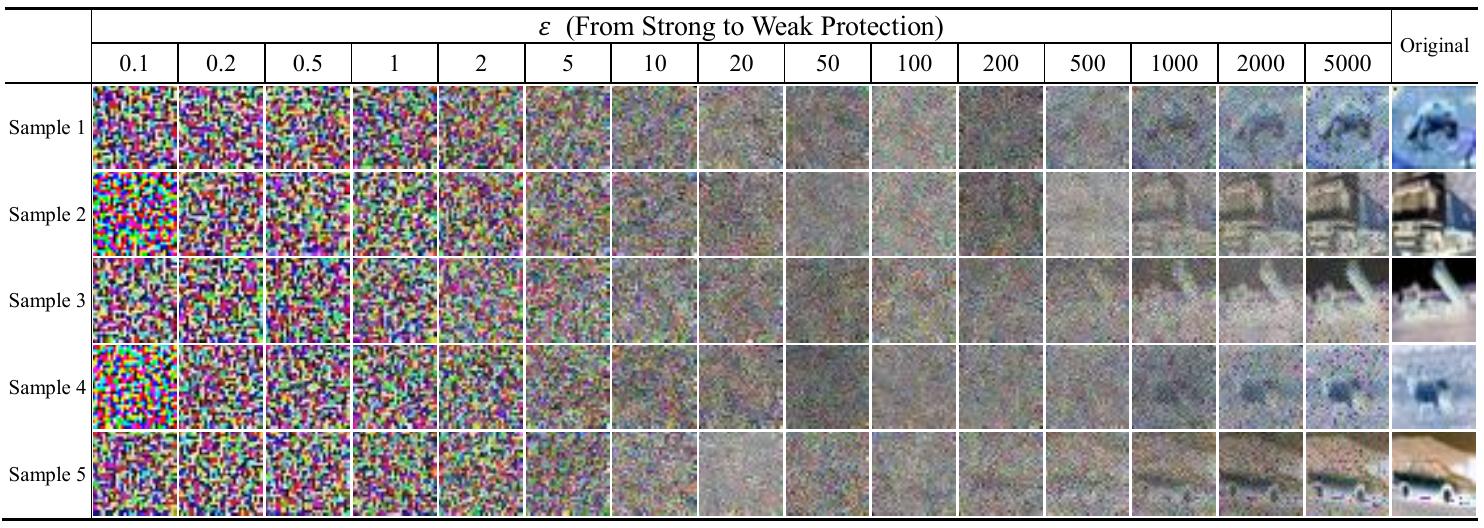}} \\ CIFAR-10 Case 2
   \end{minipage}
\caption{Visual results of applying the attack against the DP method. (Case 1 and Case 2)}
\label{fig:VisualResultsCase12}
\vspace{-15pt}
\end{figure*}

\begin{table*}[t!]
\centering
\caption{Summary of Quantitative Evaluation Results on SVHN}
\resizebox{\textwidth}{!}{
\begin{tabular}{@{}c|c|ccccccccccccccc@{}}
\toprule
SVHN&$\epsilon$&     0.1&  0.2&  0.5&  1& 2& 5& 10&   20&   50&   100&  200&  500&  1000& 2000& 5000\\ \midrule 
&  Accuracy&   9.086&   12.916&  14.319&  12.673&  90.977&  93.081&  93.479&  93.311&  93.289&  93.193&  93.129&  93.284&  93.199&  92.791&  93.072\\ 
Case 1&  MSE&  9997.76& 9518.74& 8574.61& 7726.64& 6521.06& 4947.47& 4153.4&  3796.94& 3585.67& 3550.77& 3301.97& 2593.92& 1737.37& 1394.46& 1302.65\\
&  SSIM& 0.01& 0.01& 0.01& 0.02& 0.02& 0.04& 0.07& 0.11& 0.13& 0.14& 0.16& 0.22& 0.31& 0.39& 0.49\\
&  PSNR& 27.89&   27.9& 27.9& 27.92&   27.91&   27.91&   27.89&   27.87&   27.85&   27.86&   27.88&   27.89&   28.01&   28.22&   28.47\\ \midrule 
&  Accuracy&   7.492&   7.366&   7.635&   10.336&  11.108&  85.787&  89.473&  91.364&  92.090&  91.880&  91.370&  91.963&  92.083&  91.925&  91.853\\
Case 2&  MSE&  8401.52& 8079.89& 7158.49& 6095.42& 5111.37& 4077.23& 3766.98& 3755.65& 3707.14& 3640.8&  3262.77& 3023.88& 2307.26& 1679.77& 1179.61\\
&  SSIM& 0.02& 0.02& 0.02& 0.03& 0.05& 0.09& 0.13& 0.17& 0.19& 0.19& 0.21& 0.26& 0.3&  0.37& 0.45\\
&  PSNR& 27.93&   27.91&   27.91&   27.91&   27.91&   27.88&   27.88&   27.87&   27.89&   27.87&   27.87&   27.83&   27.89&   28.02&   28.29\\ \midrule 
&  Accuracy&   11.820&  7.795&   7.338&   10.563&  12.673& 88.686& 90.397& 91.290& 91.974& 92.083& 92.097& 92.027& 92.249& 92.017& 92.031\\
Case 3&  MSE&  7140.53& 6782.47& 5795.56& 4994.62& 4239.78& 3769.36& 3696.93& 3495.58& 3649.12& 3566.92& 3989.44& 3476.29& 3112.32& 2578.59& 1781.42\\
&  SSIM& 0.02& 0.03& 0.04& 0.05& 0.09& 0.16& 0.2&  0.23& 0.24& 0.23& 0.22& 0.21& 0.25& 0.32& 0.4\\
&  PSNR& 27.9& 27.91&   27.9& 27.89&   27.88&   27.85&   27.83&   27.85&   27.89&   27.9& 27.88&   27.86&   27.86&   27.91&   28.01\\
\bottomrule
\end{tabular}}
\label{table:svhnTable}
\vspace{-5pt}
\end{table*}

\begin{table*}[t!]
\centering
\caption{Summary of Quantitative Evaluation Results on GTSRB}
\resizebox{\textwidth}{!}{
\begin{tabular}{@{}c|c|ccccccccccccccc@{}}
\toprule
GTSRB&$\epsilon$&    0.1&  0.2&  0.5&  1& 2& 5& 10&   20&   50&   100&  200&  500&  1000& 2000& 5000\\ \midrule 
&  Accuracy&   10.201&  7.654&   10.663&  10.540&  66.166&  83.544&  90.067&  91.141&  91.565&  91.287&  91.128&  91.938&  91.533&  92.236&  91.452\\
Case 1&  MSE&  12071.39&   11571.36&   10665.28&   9834.47& 8647.24& 7068.13& 6209.64& 5771.9&  5444.03& 5078.06& 4425.78& 2936.96& 1805.7&  966.69&  548.52\\
&  SSIM& 0.01& 0.01& 0.02& 0.02& 0.02& 0.03& 0.04& 0.06& 0.08& 0.12& 0.18& 0.36& 0.53& 0.68& 0.78\\
&  PSNR& 27.91&   27.89&   27.9& 27.9& 27.89&   27.89&   27.88&   27.89&   27.88&   27.85&   27.84&   27.83&   27.93&   28.18&   29.22\\ \midrule 
&  Accuracy&   10.129&  11.083&  11.381&  10.628&  14.764&  68.360&  85.811&  91.411&  92.072&  92.926&  92.936&  92.711&  92.535&  93.034&  92.522\\
Case 2&  MSE&  10188.53&   10071.2& 9580.27& 8853.6&  7773.82& 6539.38& 5858.98& 5538.74& 5379.79& 5277.89& 5039.41& 4247.2&  3224.56& 2366.67& 1343.37\\
&  SSIM& 0.01& 0.01& 0.01& 0.01& 0.02& 0.03& 0.05& 0.06& 0.08& 0.09& 0.12& 0.21& 0.33& 0.46& 0.63\\
&  PSNR& 27.91&   27.88&   27.9& 27.9& 27.91&   27.88&   27.89&   27.89&   27.87&   27.87&   27.85&   27.81&   27.8& 27.85&   27.98\\ \midrule 
&  Accuracy&   6.199& 12.089& 9.442& 7.525& 7.832& 23.742& 66.816& 85.033& 87.855& 88.025& 88.486& 88.858& 89.587& 89.756& 89.417\\
Case 3&  MSE&  10325.88&   8601.28& 7800.56& 6964.68& 6215.12& 5670.55& 5612.39& 5556.87& 5564.47& 5468.64& 5505.66& 4901.11& 3900.2&  2875.03& 1939.66\\
&  SSIM& 0.03& 0.02& 0.02& 0.03& 0.03& 0.05& 0.06& 0.07& 0.07& 0.08& 0.1&  0.16& 0.27& 0.41& 0.57\\
&  PSNR& 27.87&   27.91&   27.9& 27.9& 27.9& 27.91&   27.89&   27.88&   27.88&   27.9& 27.9& 27.87&   26.81&   27.82&   27.89\\
\bottomrule
\end{tabular}}
\label{table:gtsrbTable}
\vspace{-5pt}
\end{table*}

\begin{table*}[t!]
\centering
\caption{Summary of Quantitative Evaluation Results on STL-10}
\resizebox{\textwidth}{!}{
\begin{tabular}{@{}c|c|ccccccccccccccc@{}}
\toprule
STL-10&$\epsilon$&      0.1&  0.2&  0.5&  1& 2& 5& 10&   20&   50&   100&  200&  500&  1000& 2000& 5000\\ \midrule 
&  Accuracy&   10.105&  10.183&  10.317&  47.362&  55.837&  62.393&  64.670&  65.461&  65.693&  66.207&  66.023&  65.935&  65.857&  66.232&  65.055\\
Case 1&  MSE&  11095.13&   10570.25&   9701.28& 8717.42& 7668.91& 6058.02& 5181.42& 4764.48& 4661.07& 4352.32& 3849.4&  2714.01& 1742.79& 953.52&  602.52\\
&  SSIM& 0.01& 0.01& 0.01& 0.01& 0.02& 0.03& 0.05& 0.08& 0.1&  0.13& 0.19& 0.34& 0.49& 0.62& 0.71\\
&  PSNR& 27.88&   27.89&   27.89&   27.89&   27.91&   27.89&   27.9& 27.89&   27.89&   27.88&   27.85&   27.89&   28.02&   28.3& 29.16\\\midrule
&  Accuracy&   9.692&   10.602&  10.115&  10.099&  9.443&   46.263&  50.149&  59.110&  60.373&  63.003&  63.286&  63.707&  64.289&  64.098&  64.178\\
Case 2&  MSE&  9487.52& 9394.37& 8887.52& 8006.75& 6872.39& 5656.2&  5194.35& 4999.85& 4688.91& 4622.9&  4468.96& 3711.47& 2927.61& 2201.92& 1212.14\\
&  SSIM& 0.02& 0.02& 0.02& 0.02& 0.03& 0.04& 0.06& 0.09& 0.11& 0.12& 0.15& 0.24& 0.34& 0.45& 0.59\\
&  PSNR& 27.89&   27.88&   27.87&   27.9& 27.9& 27.88&   27.91&   27.9& 27.86&   27.87&   27.85&   27.86&   27.87&   27.91&   28.16\\\midrule
&  Accuracy&   10.177& 10.460& 9.958& 10.105& 10.622& 54.072& 57.593& 59.896& 61.983& 62.598& 62.238& 63.508& 62.621& 63.872& 62.952\\
Case 3&  MSE&  8202.77& 7868.32& 7055.42& 6097.01& 5379.97& 4930.15& 4795.49& 4763.31& 4539.6&  4760.15& 4757.22& 4331.99& 3597.64& 2583.19& 1752.03\\
&  SSIM& 0.02& 0.02& 0.02& 0.04& 0.05& 0.08& 0.1&  0.12& 0.12& 0.12& 0.14& 0.18& 0.28& 0.4&  0.54\\
&  PSNR& 27.9& 27.9& 27.9& 27.9& 27.89&   27.88&   27.91&   27.88&   27.87&   27.89&   27.91&   27.86&   27.86&   27.89&   28.02\\
\bottomrule
\end{tabular}}
\label{table:stl10Table}
\vspace{-5pt}
\end{table*}

\begin{table*}[t!]
\centering
\caption{Summary of Quantitative Evaluation Results on CIFAR-10}
\resizebox{\textwidth}{!}{
\begin{tabular}{@{}c|c|ccccccccccccccc@{}}
\toprule
CIFAR-10&$\epsilon$&    0.1&  0.2&  0.5&  1& 2& 5& 10&   20&   50&   100&  200&  500&  1000& 2000& 5000\\ \midrule 
&  Accuracy&   9.376&   9.865&   9.953&   10.256&  67.578&  74.290&  74.645&  75.424&  75.395&  75.940&  75.905&  75.756&  76.588&  75.816&  76.098\\
Case 1&  MSE&  10316.43&   9765.56& 8875.63& 7957.14& 6869.52& 5264.82& 4380.35& 4072.72& 3843.67& 3698.5&  3302.37& 2346.79& 1353.97& 795.22&  550.79\\
&  SSIM& 0.01& 0.01& 0.01& 0.01& 0.02& 0.04& 0.05& 0.08& 0.1&  0.13& 0.19& 0.33& 0.49& 0.62& 0.7\\
&  PSNR& 27.88&   27.91&   27.9& 27.9& 27.91&   27.91&   27.89&   27.87&   27.88&   27.87&   27.89&   27.9& 28.11&   28.46&   29.32\\\midrule 
&  Accuracy&   10.245&  10.027&  9.605&   10.097&  9.958&   9.985&   55.798&  60.954&  64.976&  65.707&  66.475&  65.541&  65.462&  66.286&  65.451\\
Case 2&  MSE&  8742.84& 8657.68& 8084.87& 7303.02& 6171.18& 4902.38& 4277.36& 4028.51& 4123.44& 3897.66& 3715.12& 3113.49& 2576.85& 1750.36& 1002\\
&  SSIM& 0.02& 0.02& 0.02& 0.02& 0.03& 0.04& 0.07& 0.09& 0.11& 0.12& 0.15& 0.23& 0.33& 0.45& 0.58\\
&  PSNR& 27.9& 27.9& 27.91&   27.9& 27.9& 27.89&   27.87&   27.89&   27.87&   27.87&   27.87&   27.87&   27.87&   27.97&   28.27\\\midrule 
&  Accuracy&   9.621& 9.882& 9.985& 10.058& 10.221& 10.339& 11.006& 51.949& 61.840& 67.829& 69.755& 69.610& 72.114& 70.960& 71.559\\
Case 3&  MSE&  7518.53& 7298.3&  6497.63& 5598.26& 4845.48& 4179.38& 4051.71& 4013.06& 3836.35& 3954.54& 3921.73& 3689.95& 3241.79& 2475.87& 1734.69\\
&  SSIM& 0.02& 0.02& 0.03& 0.03& 0.05& 0.08& 0.11& 0.12& 0.13& 0.13& 0.13& 0.16& 0.23& 0.36& 0.5\\
&  PSNR& 27.91&   27.89&   27.9& 27.9& 27.89&   27.87&   27.86&   27.87&   27.9& 27.89&   27.87&   27.88&   27.89&   27.94&   28\\
\bottomrule
\end{tabular}}
\label{table:cifar10Table}
\end{table*}

\section{More Visual and Quantitative Evaluation Results}
\label{sec:AppendotherResulsts}
Fig. \ref{fig:VisualResultsCase12} show some visual evaluation results on Case 1 and Case 2 in datasets (SVHN, GTSRB, STL-10, CIFAR-10) regarding the protection levels of the DP method against the data reconstruction attack.
We can see that the reconstruction attack is not effective even for smaller $\epsilon$ value as the local part model layer increases. It is observed that even at $\epsilon$ = 1000 in Case 1, the reconstructed images reveal meaningful visual information of the original images, in Case 2, the reconstructed images, the reconstructed images almost reveal no meaningful information of the original images.

Tables \ref{table:svhnTable}-\ref{table:cifar10Table} provide the quantitative evaluation results in terms of accuracy, MSE, SSIM, and PSNR.
Note that the accuracy results were plotted in Fig. \ref{fig:svhn_acc}, Fig. \ref{fig:gtsrb_acc}, Fig. \ref{fig:stl10_acc}, and Fig. \ref{fig:cifar10_acc}.
And the MSE, SSIM, and PSNR results were plotted in Fig. \ref{fig:quantified_results_svhn}, Fig. \ref{fig:quantified_results_gtsrb}, Fig. \ref{fig:quantified_results_stl10}, and Fig. \ref{fig:quantified_results_cifar10}.
We provide the exact figures here to facilitate the observations.

\bibliographystyle{spmpsci}     
\bibliography{references.bib}

\begin{thebibliography}{10}
\providecommand{\url}[1]{#1}
\csname url@samestyle\endcsname
\providecommand{\newblock}{\relax}
\providecommand{\bibinfo}[2]{#2}
\providecommand{\BIBentrySTDinterwordspacing}{\spaceskip=0pt\relax}
\providecommand{\BIBentryALTinterwordstretchfactor}{4}
\providecommand{\BIBentryALTinterwordspacing}{\spaceskip=\fontdimen2\font plus
\BIBentryALTinterwordstretchfactor\fontdimen3\font minus
  \fontdimen4\font\relax}
\providecommand{\BIBforeignlanguage}[2]{{%
\expandafter\ifx\csname l@#1\endcsname\relax
\typeout{** WARNING: IEEEtran.bst: No hyphenation pattern has been}%
\typeout{** loaded for the language `#1'. Using the pattern for}%
\typeout{** the default language instead.}%
\else
\language=\csname l@#1\endcsname
\fi
#2}}
\providecommand{\BIBdecl}{\relax}
\BIBdecl

\bibitem{YaoHZZA17}
S.~Yao, S.~Hu, Y.~Zhao, A.~Zhang, and T.~F. Abdelzaher, ``Deepsense: {A}
  unified deep learning framework for time-series mobile sensing data
  processing,'' in \emph{Proc. of WWW}, 2017.

\bibitem{RaduTBLMMK17}
V.~Radu, C.~Tong, S.~Bhattacharya, N.~D. Lane, C.~Mascolo, M.~K. Marina, and
  F.~Kawsar, ``Multimodal deep learning for activity and context recognition,''
  \emph{Proc. {ACM} Interact. Mob. Wearable Ubiquitous Technol.}, vol.~1,
  no.~4, pp. 157:1--157:27, 2017.

\bibitem{YaoZSZZLA17}
S.~Yao, Y.~Zhao, H.~Shao, A.~Zhang, C.~Zhang, S.~Li, and T.~F. Abdelzaher,
  ``Rdeepsense: Reliable deep mobile computing models with uncertainty
  estimations,'' \emph{Proc. {ACM} Interact. Mob. Wearable Ubiquitous
  Technol.}, vol.~1, no.~4, pp. 173:1--173:26, 2017.

\bibitem{YaoZSZZHLLSA18}
S.~Yao, Y.~Zhao, H.~Shao, C.~Zhang, A.~Zhang, S.~Hu, D.~Liu, S.~Liu, L.~Su, and
  T.~F. Abdelzaher, ``Sensegan: Enabling deep learning for internet of things
  with a semi-supervised framework,'' \emph{Proc. {ACM} Interact. Mob. Wearable
  Ubiquitous Technol.}, vol.~2, no.~3, pp. 144:1--144:21, 2018.

\bibitem{yao2018deep}
S.~Yao, Y.~Zhao, A.~Zhang, S.~Hu, H.~Shao, C.~Zhang, L.~Su, and T.~Abdelzaher,
  ``Deep learning for the internet of things,'' \emph{Computer}, vol.~51,
  no.~5, pp. 32--41, 2018.

\bibitem{YaoZSLLSA18}
S.~Yao, Y.~Zhao, H.~Shao, S.~Liu, D.~Liu, L.~Su, and T.~F. Abdelzaher,
  ``Fastdeepiot: Towards understanding and optimizing neural network execution
  time on mobile and embedded devices,'' in \emph{Proc. of ACM SenSys}, 2018.

\bibitem{Teerapittayanon17}
S.~Teerapittayanon, B.~McDanel, and H.~T. Kung, ``Distributed deep neural
  networks over the cloud, the edge and end devices,'' in \emph{Proc. of IEEE
  ICDCS}, 2017.

\bibitem{KoNAM18}
J.~H. Ko, T.~Na, M.~F. Amir, and S.~Mukhopadhyay, ``Edge-host partitioning of
  deep neural networks with feature space encoding for resource-constrained
  internet-of-things platforms,'' in \emph{Proc. of {IEEE} International
  Conference on Advanced Video and Signal Based Surveillance}, 2018.

\bibitem{WangZBZCY18}
J.~Wang, J.~Zhang, W.~Bao, X.~Zhu, B.~Cao, and P.~S. Yu, ``Not just privacy:
  Improving performance of private deep learning in mobile cloud,'' in
  \emph{Proc. of KDD}, 2018.

\bibitem{HeZL19}
Z.~He, T.~Zhang, and R.~B. Lee, ``Model inversion attacks against collaborative
  inference,'' in \emph{Proc. of ACSAC}, 2019.

\bibitem{Dwork06}
C.~Dwork, ``Differential privacy,'' in \emph{Proc. of {ICALP}}, 2006.

\bibitem{DworkMNS06}
C.~Dwork, F.~McSherry, K.~Nissim, and A.~D. Smith, ``Calibrating noise to
  sensitivity in private data analysis,'' in \emph{Proc. of TCC}, 2006.

\bibitem{BaiLLYJX22}
J.~Bai, Y.~Li, J.~Li, X.~Yang, Y.~Jiang, and S.~Xia, ``Multinomial random
  forest,'' \emph{Pattern Recognition}, vol. 122, p. 108331, 2022.

\bibitem{svhn}
Y.~Netzer, T.~Wang, A.~Coates, A.~Bissacco, B.~Wu, and A.~Y. Ng, ``Reading
  digits in natural images with unsupervised feature learning,'' in \emph{ICLR
  AI for social good workshop}, 2011.

\bibitem{gtsrb}
J.~Stallkamp, M.~Schlipsing, J.~Salmen, and C.~Igel, ``Man vs. computer:
  Benchmarking machine learning algorithms for traffic sign recognition,''
  \emph{Neural Networks}, vol.~32, pp. 323--332, 2012.

\bibitem{cifar10}
A.~Krizhevsky, ``Learning multiple layers of features from tiny images,'' Tech.
  Rep., 2009.

\bibitem{stl10}
A.~Coates, A.~Y. Ng, and H.~Lee, ``An analysis of single-layer networks in
  unsupervised feature learning,'' in \emph{Proc. of {AISTATS}}, 2011.

\bibitem{Jayaraman019}
B.~Jayaraman and D.~Evans, ``Evaluating differentially private machine learning
  in practice,'' in \emph{Proc. of USENIX Security}, 2019.

\bibitem{wang2004image}
Z.~Wang, A.~C. Bovik, H.~R. Sheikh, and E.~P. Simoncelli, ``Image quality
  assessment: from error visibility to structural similarity,'' \emph{IEEE
  transactions on image processing}, vol.~13, no.~4, pp. 600--612, 2004.

\bibitem{dosovitskiy2014discriminative}
A.~Dosovitskiy, J.~T. Springenberg, M.~Riedmiller, and T.~Brox,
  ``Discriminative unsupervised feature learning with convolutional neural
  networks,'' in \emph{Proc. of NeurlPS}, 2014, pp. 766--774.

\bibitem{huang2020privacy}
H.~Huang, D.~Zhang, F.~Xiao, K.~Wang, J.~Gu, and R.~Wang, ``Privacy-preserving
  approach pbcn in social network with differential privacy,'' \emph{IEEE
  Transactions on Network and Service Management}, vol.~17, no.~2, pp.
  931--945, 2020.

\bibitem{nguyen2020privacy}
D.~C. Nguyen, P.~N. Pathirana, M.~Ding, and A.~Seneviratne, ``Privacy-preserved
  task offloading in mobile blockchain with deep reinforcement learning,''
  \emph{IEEE Transactions on Network and Service Management}, vol.~17, no.~4,
  pp. 2536--2549, 2020.

\bibitem{andreoletti2020privacy}
D.~Andreoletti, T.~Velichkova, G.~Verticale, M.~Tornatore, and S.~Giordano, ``A
  privacy-preserving reinforcement learning algorithm for multi-domain virtual
  network embedding,'' \emph{IEEE Transactions on Network and Service
  Management}, vol.~17, no.~4, pp. 2291--2304, 2020.

\bibitem{dong2021network}
S.~Dong, Y.~Xia, and T.~Peng, ``Network abnormal traffic detection model based
  on semi-supervised deep reinforcement learning,'' \emph{IEEE Transactions on
  Network and Service Management}, 2021.

\bibitem{khan2021socially}
L.~U. Khan, Z.~Han, D.~Niyato, and C.~S. Hong,
  ``Socially-aware-clustering-enabled federated learning for edge networks,''
  \emph{IEEE Transactions on Network and Service Management}, 2021.

\bibitem{zhang2016fakemask}
L.~Zhang, Z.~Cai, and X.~Wang, ``Fakemask: A novel privacy preserving approach
  for smartphones,'' \emph{IEEE Transactions on Network and Service
  Management}, vol.~13, no.~2, pp. 335--348, 2016.

\bibitem{subramanya2021centralized}
T.~Subramanya and R.~Riggio, ``Centralized and federated learning for
  predictive vnf autoscaling in multi-domain 5g networks and beyond,''
  \emph{IEEE Transactions on Network and Service Management}, vol.~18, no.~1,
  pp. 63--78, 2021.

\bibitem{ding2019extended}
W.~Ding, R.~Hu, Z.~Yan, X.~Qian, R.~H. Deng, L.~T. Yang, and M.~Dong, ``An
  extended framework of privacy-preserving computation with flexible access
  control,'' \emph{IEEE Transactions on Network and Service Management},
  vol.~17, no.~2, pp. 918--930, 2019.

\bibitem{groleat2012distributed}
T.~Groleat and H.~Pouyllau, ``Distributed learning algorithms for inter-nsp sla
  negotiation management,'' \emph{IEEE Transactions on Network and Service
  Management}, vol.~9, no.~4, pp. 433--445, 2012.

\bibitem{ZhengLLYYW22}
Y.~Zheng, S.~Lai, Y.~Liu, X.~Yuan, X.~Yi, and C.~Wang, ``Aggregation service
  for federated learning: An efficient, secure, and more resilient
  realization,'' \emph{{IEEE} Transactions on Dependable and Secure Computing},
  2022, doi: \url{10.1109/TDSC.2022.3146448}.

\bibitem{ZhuLLYXL22}
L.~Zhu, X.~Liu, Y.~Li, X.~Yang, S.~Xia, and R.~Lu, ``A fine-grained
  differentially private federated learning against leakage from gradients,''
  \emph{{IEEE} Internet of Things Journal}, vol.~9, no.~13, pp.
  11\,500--11\,512, 2022.

\bibitem{ZhengDTWZ21}
Y.~Zheng, H.~Duan, X.~Tang, C.~Wang, and J.~Zhou, ``Denoising in the dark:
  Privacy-preserving deep neural network-based image denoising,'' \emph{{IEEE}
  Transactions on Dependable and Secure Computing}, vol.~18, no.~3, pp.
  1261--1275, 2021.

\bibitem{LiuZYY21}
X.~Liu, Y.~Zheng, X.~Yuan, and X.~Yi, ``Medisc: Towards secure and lightweight
  deep learning as a medical diagnostic service,'' in \emph{Proc. of ESORICS},
  2021.

\bibitem{RahmanRLM18}
M.~A. Rahman, T.~Rahman, R.~Lagani{\`{e}}re, and N.~Mohammed, ``Membership
  inference attack against differentially private deep learning model,''
  \emph{Transactions on Data Privacy}, vol.~11, no.~1, pp. 61--79, 2018.

\bibitem{AbadiCGMMT016}
M.~Abadi, A.~Chu, I.~J. Goodfellow, H.~B. McMahan, I.~Mironov, K.~Talwar, and
  L.~Zhang, ``Deep learning with differential privacy,'' in \emph{Proc. of ACM
  CCS}, 2016.

\bibitem{abs-1912-11328}
D.~Bernau, P.~Grassal, J.~Robl, and F.~Kerschbaum, ``Assessing differentially
  private deep learning with membership inference,'' \emph{CoRR}, vol.
  abs/1912.11328, 2019.

\end{thebibliography}

\end{document}